\documentstyle[12pt,amsmath,amssymb,epsf,epic]{article}

\numberwithin{equation}{section}

\newcommand{\un}{{\mathbb I}}
\newcommand{\ra}{\rightarrow}
\newcommand{\tr}{\mbox{Tr}}

\newcommand{\spr}{\mbox{Spr}\,} 
\newcommand{\Bra}{\big\langle} 
\newcommand{\bra}{\langle} 
\newcommand{\Ket}{\big\rangle}
\newcommand{\ket}{\rangle}

\newcommand{\E}{{\mathbb E}}
\newcommand{\D}{{\mathbb D}}

\newcommand{\be}{\begin{equation}}
\newcommand{\ee}{\end{equation}}
\newcommand{\bea}{\begin{eqnarray}}
\newcommand{\eea}{\end{eqnarray}}

\newcommand{\ffi}{\varphi}

\newcommand{\ep}{\hfill  {\vrule height 10pt width 8pt depth 0pt}}

\newcommand{\grintl}{[\kern-.18em [}
\newcommand{\grintr}{]\kern-.18em ]}

\newcounter{resultcounter}[section]

\newtheorem{thm}[resultcounter]{Theorem}
\newtheorem{lem}[resultcounter]{Lemma}
\newtheorem{prop}[resultcounter]{Proposition}
\newtheorem{cor}[resultcounter]{Corollary}
\newtheorem{definition}[resultcounter]{Definition}
\newtheorem{rem}[resultcounter]{Remark}
\newtheorem{rems}[resultcounter]{Remarks}

\def\bed{\begin{definition}}
\def\eed{\end{definition}}

\def\proof{\noindent{\bf Proof.}\ \ }

 \def\cB{{\cal B}} 
\def\cD{{\cal D}}  \def\cF{{\cal F}}
\def\cG{{\cal G}} \def\cH{{\cal H}} \def\cI{{\cal I}}
 \def\cK{{\cal K}} 
\def\cM{{\cal M}} \def\cN{{\cal N}} 
\def\cP{{\cal P}}  
 \def\cT{{\cal T}}

\newcommand{\R}{{\mathbb R}}
\newcommand{\N}{{\mathbb N}}
\newcommand{\Q}{{\mathbb Q}}
\newcommand{\C}{{\mathbb C}}
\newcommand{\Z}{{\mathbb Z}}
\renewcommand{\E}{{\mathbb E}}
\renewcommand{\P}{{\mathbb P}}
\newcommand{\I}{{\mathbb I}}
\newcommand{\T}{{\mathbb T}}

\def\proof{\noindent{\bf Proof:}\ \ }

\begin{document}
\title{Correlated Markov Quantum Walks}
 \author{ Eman Hamza\footnote{Department of Physics, Faculty of Science, Cairo University, Cairo 12613, Egypt}\and Alain Joye\footnote{ UJF-Grenoble 1, CNRS Institut Fourier UMR 5582, Grenoble, 38402, France} \footnote{Partially supported by the Agence Nationale de la Recherche, grant ANR-09-BLAN-0098-01}}

\date{ }

\maketitle
\vspace{-1cm}

\thispagestyle{empty}
\setcounter{page}{1}
\setcounter{section}{1}

\setcounter{section}{0}

\abstract{

We consider the discrete time unitary dynamics given by a quantum walk on $\Z^d$ performed by a particle with internal degree of freedom, called coin state,  according to the following iterated rule: a unitary update of the coin state takes place, followed by a shift on the lattice, conditioned on the coin state of the particle. 
We study the large time behavior of the quantum mechanical probability distribution of the position observable in $\Z^d$ for random updates of the coin states of the following form. The random sequences of unitary updates are given by a site dependent function of a Markov chain in time, with the following properties: on each site, they share the same stationnary Markovian distribution and, for each fixed time, they form a deterministic  periodic pattern on the lattice. 

We prove a Feynman-Kac formula to express the characteristic function of the averaged distribution over the randomness at time $n$ in terms of the $n^{\mbox{th}}$ power of an operator $M$. By analyzing  the spectrum of $M$, we show that  this distribution posesses a drift proportional to the time and its centered counterpart displays a diffusive behavior with a diffusion matrix we compute. Moderate and large deviations principles are also proven to hold for the averaged distribution and the limit of the suitably rescaled corresponding characteristic function is shown to satisfy a diffusion equation. 

An example of random updates for which the analysis of the distribution can be performed without averaging is worked out. The random distribution displays a deterministic drift proportional to time and its centered counterpart gives rise to a random diffusion matrix whose law we compute. We complete the picture by presenting an uncorrelated example. 
}

\thispagestyle{empty}
\setcounter{page}{1}
\setcounter{section}{1}

\setcounter{section}{0}

\section{Introduction}

Quantum walks are simple models of discrete time quantum evolution taking place on a $d$-dimensional lattice whose implementation yields a unitary discrete dynamical system on a Hilbert space. The dynamics describes the motion of a quantum particle with internal degree of freedom on an infinite $d$-dimensional lattice according to the following rules. The one-step motion consists in an update of the internal degree of freedom by means of a unitary transform in the relevant part of the Hilbert space, followed by a finite range shift on the lattice, conditioned on the internal degree of freedom of the particle. Due to their similarity with classical random walks on a lattice, quantum walks constructed this way are often considered as their quantum analogs. In this context, the space of the internal degree of freedom is called {\it coin space}, the degree of freedom is the {\it coin state} and the unitary operators performing the update are {\it coin matrices}.

Quantum walks have become quite popular in the quantum computing community in the recent years, due to the role they play in computer science, and in particular for quantum search algorithms. See for example \cite{M},  \cite{AAKV},  \cite{Ke}, \cite{Ko}, \cite{SKW}, \cite{AKR}, \cite{MNRS} and in the review \cite{S}. 
Also, quantum walks are used as effective dynamics of quantum systems in certain asymptotic regimes; see {\em e. g.} \cite{cc}, \cite{ADZ}, \cite{M}, \cite{lv},  \cite{bb}, \cite{rhk}, for a few models of this type, and \cite{ade}, \cite{BHJ}, \cite{dOS}, \cite{HJS}, \cite{abj} for their mathematical analysis.
Moreover, quantum walk dynamics have been shown to describe experimental reality for systems of cold atoms trapped in suitably monitored optical lattices  \cite{Ketal}, and ions caught in monitored Paul traps \cite{Zetal}.

\medskip

The literature contains several variants of the quantum dynamics on a lattice as described above, which may include decoherence effects and/or more general graph,  see {\em e.g.} the reviews and papers \cite{AAKV}, \cite{Ke}, \cite{avww}, \cite{apss}.
In this work, we consider the case where the evolution of the walker is unitary, and where the underlying lattice is $\Z^d$ with coin space of dimension $2d$, which is, in a sense, the closest to the classical random walk. 

We are interested in the long time behavior of quantum mechanical expectation values of observables that are non-trivial on the lattice only, {\em i.e.} that do not depend on the internal degree of freedom of the quantum walker. Equivalently, this amounts to studying a family of random vectors $X_n$ on the lattice $\Z^d$, indexed by the discrete time variable, with probability laws $\P(X_n=k)=W_k(n)$ defined by the prescriptions of quantum mechanics. The initial state of the quantum walker is described by a density matrix.
\medskip

As is well known, when the unitary update of the coin variable is performed at each time step by means of the same coin matrix, this leads to a ballistic behavior of the expectation of the position variable characterized by $\E_{W(n)}(X_n)\simeq n V$ when $n$ is large, for some vector $V$, and by fluctuations of the centered random variable $X_n-nV$ of order $n$, 
see {\em e.g.} \cite{Ko}.

The case where the coin matrices used to update the coin variable depend on the time step in a random fashion, a situation of {\it temporal disorder}, is dealt with 
in \cite{J1}, see also \cite{avww}. All coin variables are updated simultaneously and in the same way, independently of the position on the lattice. This yields a random distribution $W_\cdot^\omega(n)$, corresponding to the random variable $X_n^\omega$ which, once centered and averaged over the disorder, displays a diffusive behavior in the long time limit.

If the coin matrices depend on the site of the lattice $\Z^d$ but not on time, {\em i.e.} a case of {\em spatial disorder}, one expects dynamical localization, characterized by finite values of all moments, uniformly bounded in time $n$, and for (almost) all realizations. In dimension $d=1$, this was proven in \cite{jm} for certain sets of random coin matrices, which were further generalized in \cite{asw}. See also  \cite{Ko1}, \cite{sk}  for related aspects. The higher dimensional case is open.

\medskip

The situation addressed here is that of correlated {\em spatio-temporal disorder}. We consider random coin matrices which depend both on time and space in the following way: The random coin matrix at site $x\in \Z^d$ and time $n\in\N$ is given by
$
C^\omega_n(x)=\sigma_x(\omega(n)),
$
where $\{\omega(j)\}_{j\in\N}$ is a temporal stationary Markov chain on a finite set $\Omega$ of unitary matrices on $\C^{2d}$, 
and $\Z^d\ni x\ra \sigma_x$ is a given representation of $\Z^d$ in terms of measure invariant bijections on $\Omega$. In particular, $\sigma_0=$Id, the identity on $\Omega$, and $\Gamma=\{y\in \Z^d\ \mbox{s.t.}\ \sigma_y=\mbox{Id}\}$ forms a periodic sub-lattice  of $\Z^d$. Therefore, at each site $x\in\Z^d$, the sequence $\{C^\omega_j(x)\}_{j\in\N}$ is Markovian with a distribution independent of $x$, and at each time $n\in \N$, the set $\{C^\omega_n(x), \ x\in \Z^d\}$ is $\Gamma$-periodic. 
This is a natural generalization of the case studied in \cite{J1} which displays a deterministic non trivial periodic structure in the spatial patterns of random coin matrices at each time step.

This setup is an analog of the one addressed in \cite{p}, \cite{ks}, \cite{hks}, where the dynamics is generated by a quantum Hamiltonian with a time dependent potential generated by a random process. For quantum walks, the role of the random time dependent potential is played by the random coin operators whereas the role of the deterministic kinetic energy is played by the shift. 

\medskip

We address the problem by an analysis of the large $n$ behavior of the characteristic function of the distribution $w_\cdot(n)$, $\Phi_n(y)=\E_{w(n)}(e^{iyX_n})$, where $w_\cdot(n)=\E(W_\cdot^\omega(n))$ is the averaged quantum mechanical distribution on $\Z^d$, with initial condition $\rho_0$, a density matrix on $l^2(\Z^d)\otimes \C^{2d}$. 
By adapting the strategy of \cite{ks}, \cite{hks}, inspired by \cite{p}, to our discrete time unitary setup, we first establish a {\em Feynman-Kac} type formula to express $w_\cdot(n)$ in terms of (some matrix element) of the $n^{\mbox{th}}$ power of a contraction operator $M$ acting on an extended Hilbert space which involves a space of (density) matrices and the probability space of coin matrices. Then, we analyze the spectral properties of $M$, making use of the periodicity and invariance properties of $\sigma_x$ which yield a fiber decomposition of a generalized Fourier transform of $M$. In turn, this allows us to provide a detailed description of the large $n$ behavior of the characteristic function $\Phi_n(y)$ in the diffusive regime  $y\ra y/\sqrt n$, and at $y$ fixed, in terms of the spectral data of $M$ and their perturbative behavior. 

The foregoing is the main technical result of the paper, from which several consequences can be drawn, by arguments similar to those used in \cite{J1}.
Under natural assumptions on the spectrum of $M$, the averaged distribution $w_\cdot(n)$ displays a diffusive behavior characterized by the following data: a deterministic drift vector $\overline r\in \R^d$ and a diffusion matrix $\D$, which we compute, such that, for $n$ large and $ i,j =1,2, \cdots, d$,
$$\E_{w(n)}(X_n)\simeq n \overline r, \ \ \E_{w(n)}((X_n-n\overline r)_i(X_n-n\overline r)_j)\simeq n\D_{ij}.$$ Moreover, we get convergence of the properly rescaled characteristic function of $X_n-n\overline r$, $e^{-i[tn]\overline ry/\sqrt{n}}\Phi_{[tn]}(y/\sqrt{n})$, to the Fourier transform of superpositions of solutions to a diffusion equation of the form $
\int_{\T^d}e^{-\frac t2\bra y | \D(p) y\ket}{dv}/{(2\pi)^d}
$
with diffusion matrix $\D(p)$, $p\in \T^d$, the $d$-dimensional torus. Also, we get moderate deviations results of the type  
 \be
\P(X_n-n\overline r\in n^{(\alpha +1)/2}\, \Gamma)\simeq e^{-n^\alpha \inf_{x\in {\Gamma}}\Lambda^*(x)} \ \ \mbox{as $n\ra\infty$,}
\ee
 for any set $\Gamma\in \R^d$, any  $0<\alpha<1$, 
with some {\it rate function} $\Lambda^*:\R^d\ra[0,\infty]$ we determine. Finally, we improve on \cite{J1} by establishing {\it large deviations} results for sets in a certain neighborhood of the origin, under stronger hypotheses. Informally, there exists an open ball $B$ centered at the origin such that for all sets $\Gamma\in B\cap\R^d$,
\be
\P(X_n-n\overline r\in n\, \Gamma)\simeq e^{-n \inf_{x\in {\Gamma}}\overline\Lambda^*(x)} \ \ \mbox{as $n\ra\infty$,}
\ee
 where $\overline\Lambda^*:\R^d\ra[0,\infty]$ is another {\it rate function} we determine. By Bryc's argument, \cite{b}, a central limit theorem for $X_n$ holds under the same conditions.

To complete the picture, we work out an example introduced in \cite{J1} where the distribution of coin matrices is supported on the set of unitary permutation matrices.   This case allows us to analyze of the random distribution $W^\omega(n)$, without averaging over the disorder. We show that under our hypotheses, in this case $W^\omega(n)$ coincide with the distribution of a classical walk on the lattice, with increments being neither stationary, nor Markovian. Nevertheless, we can apply spectral methods as well to study the long times asymptotics of the corresponding random characteristic function, which allows us to get the the existence of a random diffusion matrix $\D^\omega$ such that
$$
\E_{W^\omega(n)}((X_n^\omega-n\overline r)_i(X_n^\omega-n\overline r)_j)\simeq n\D^\omega_{ij}, \ i,j =1,2, \cdots, d,
$$
and whose matrix elements $\D^\omega_{ij}$ are distributed according to the law of $X_i^\omega X_j^\omega$, where the vector $X^\omega$ is distributed according to $\cN(0,\Sigma)$, for a matrix $\Sigma$ we determine.

\medskip

We also consider the completely decorrelated case where the coin matrices at each sites are i.i.d, {\em i.e.} a situation where no spatial structure is present in the pattern of coin matrices.

\medskip

\medskip

{\bf Acknowledgements} E.H. wishes to thank the CNRS and the Institut Fourier for support in the Fall of 2010, where this work was initiated, and A.J. wishes to thank the CRM for support in July 2011, where part of this work was done.

\section{General Setup}\label{dets}

Let $\cH=\C^{2d}\otimes l^2(\Z^{d})$ be the Hilbert space of the quantum walker in $\Z^d$ with $2d$ internal degrees of freedom. We denote the canonical basis of $\C^{2d}$ by $\{|\tau\ket \}_{\tau\in I^d_\pm}$, where $I_\pm=\{\pm 1, \pm 2, \dots, \pm d\}$, so that the orthogonal projectors on the basis vectors are noted  $P_\tau=|\tau\ket\bra \tau|$, $\tau \in I_\pm$.  We shall
denote the canonical basis of $l^2(\Z^{d})$ by $\{|x\ket\}_{x\in\Z^d}$, or by $\{\delta_x\}_{x\in\Z^d}$. We shall write for a vector $\psi\in \cH$,  $\psi=\sum_{x\in\Z^d}\psi(x)|x\ket$, where $\psi(x)=\bra x|\psi\ket\in \C^{2d}$ and $\sum_{x\in\Z^d}\|\psi(x)\|_{\C^{2d}}^2=\|\psi\|^2<\infty$. We shall abuse notations by using the same symbols $\bra \cdot | \cdot\ket$ for scalar products and corresponding ``bra" and ``ket" vectors on $\cH$, $\C^{2d}$ and $l^2(\Z^d)$, the context allowing us to determine which spaces we are talking about. Also, we will  often drop the subscript ${\C^{2d}}$ of the norm.

\medskip

A {\it coin matrix} acting on the internal degrees of freedom, or {\it coin state}, is a unitary matrix $C\in M_{2d}(\C)$ and a {\it jump function} is a function $r : I_\pm\ra\Z^d$. The shift $S$ is defined on $\cH$ by
\bea
S
&=&\sum_{x\in \Z^d}\sum_{\tau\in I_\pm} P_\tau\otimes |x+r(\tau)\ket\bra x|.
\eea
By construction, a walker at site $y$ with internal degree of freedom $\tau$ represented by the vector $|\tau\ket\otimes |y\ket\in \cH$ is just sent by $S$ to one of the neighboring sites depending on $\tau$ determined by the jump function $r(\tau)$
\be
S\ |\tau\ket\otimes |y\ket=|\tau\ket\otimes |y+r(\tau)\ket. 
\ee
The composition by $C(y)\otimes \I$, where the coin matrix $C(y)$ is allowed to depend on the site $y$, reshuffles or updates the coin state so that the pieces of the wave function corresponding to different internal states are shifted to different directions, depending on the internal state. The corresponding one step unitary evolution $U$  of the walker on $\cH=\C^{2d}\otimes l^2(\Z^{d})$ is given by 
\be\label{defu}
U=\sum_{x\in \Z^d}\sum_{\tau\in I_\pm} P_\tau C(x)\otimes |x+r(\tau)\ket\bra x|.
\ee

Given a set of $n>0$ site-dependent unitary coin matrices $C_k(x)\in M_{2d}(\C)$, $k=1,\cdots, n$ and $x\in \Z^d$, we construct an evolution operator $U(n,0)$ from time $0$ to time $n$, characterized at time $k$ by $U_k$ defined in (\ref{defu}) with $\{C_k(x)\}_{x\in\Z^d}$ via
\be
U(n,0)=U_nU_{n-1}\cdots U_1.
\ee
 
\medskip

Let $f:\Z^d\ra \C$ and define the multiplication operator $F:D(F)\ra \cH$ on its domain $D(F)\subset\cH$ by $(F\psi)(x)=f(x)\psi(x)$, $\forall  x\in\Z^d$, where $\psi\in D(F)$ is equivalent to $\sum_{x\in \Z^d}|f(x)|^2\|\psi(x)\|_{\C^d}^2<\infty$. Note that $F$ acts trivially on the coin state. 

When $f$ is real valued, $F$ is self-adjoint and will be called a {\it lattice observable}.
\medskip

In particular, consider a walker characterized at time zero by the normalized vector $\psi_0=\ffi_0\otimes |0\ket$, {\em i.e.} which sits on site $0$ with coin state $\ffi_0$. The quantum mechanical expectation value of a lattice observable $F$ at time $n$ is given by $\bra F \ket_{\psi_0}(n)=\bra \psi_0|U(n,0)^* F U(n,0)\psi_0\ket$.

 A straightforward computation yields the following expression for the corresponding discrete evolution from time zero to time $n$
\begin{lem}\label{expn} With the notations above, 
\bea\label{defUn}
U(n,0)&=&
\sum_{x\in \Z^d}\sum_{k\in \Z^d}J^x_k(n)\otimes |x+k\ket\bra x|,
\eea
where
\be\label{defJ}
J^x_k(n)=\sum_{\tau_1, \tau_2, \dots, \tau_n\in {I_\pm}^n\atop \sum_{s=1}^nr(\tau_s)=k} P_{\tau_n}C_{n}\Big(x+\sum_{s=1}^{n-1}r(\tau_s)\Big)P_{\tau_{n-1}}C_{n-1}\Big(\small{x+\sum_{s=1}^{n-2}r(\tau_s)}\Big)\cdots P_{\tau_1}C_{1}(x)\in M_{2d}(\C)
\ee
and $J^x_k(n)=0$, if $\sum_{s=1}^nr(\tau_s)\neq k$.
Moreover, for any lattice observable $F$, and any normalized vector $\psi_0=\ffi_0\otimes |0\ket$, 
\bea\label{rvlat}\nonumber
\bra F \ket_{\psi_0}(n)&=&\bra \psi_0|U^*(n,0)F U(n,0)\psi_0\ket=\sum_{k\in\Z^d}f(k)\bra\ffi_0|J^0_k(n)^*J^0_k(n)\ffi_0\ket\\
&\equiv&\sum_{k\in\Z^d}f(k)W_k(n),
\eea
where $W_k(n)=\|J^0_k(n)\ffi_0\|^2_{\C^{2d}}$ satisfy
\be\label{norm}
\sum_{k\in\Z^d}W_k(n)=\sum_{k\in\Z^d}\|J^0_k(n)\ffi_0\|^2_{\C^{2d}}=\|\psi_0\|^2_{\cH}=1.
\ee
\end{lem}
\begin{rem}\label{defx}
We view the non-negative quantities $\{W_k(n)\}_{n\in\N^*}$ as the probability distributions of a sequence of $\Z^d$-valued random variables $\{X_n\}_{n\in\N^*}$ with 
\be
\mbox{\em Prob}(X_n=k)=W_k(n)= \bra\psi_0 | U(n,0)^* (\I\otimes |k\ket\bra k|) U(n,0)\psi_0\ket=\|J^0_k(n)\ffi_0\|^2_{\C^{2d}},
\ee
in keeping with (\ref{rvlat}). In particular,  $\bra F \ket_{\psi_0}(n)=\E_{W_k(n)}(f(X_n))$. We shall use freely both notations.

\end{rem}
\begin{rem}
\medskip 
All sums over $k\in\Z^k$ are finite since $J^x_k(n)=0$ if $\max_{j=1,\dots,d}|k_j|>\rho n$, for some $\rho>0$ independent of $x\in\Z^d$, since the jump functions have finite range.

\end{rem}
We are particularly interested in the long time behavior, $n >\hspace{-4pt}> 1$,  of $\bra X^2\ket_{\psi_0}(n)$, the expectation of the observable $X^2$ corresponding to the function $f(x)=x^2$ on $\Z^d$ with initial condition $\psi_0$. Or, in other words, in the second moments of the distributions $\{W_k(n)\}_{n\in\N^*}$.

\medskip

Let us proceed by expressing the probabilities $W_k(n)$ in terms of the $C_k$'s, $k=1,\dots, n$. We need to introduce some more notations. Let $I_n(k)=\{\tau_1,\cdots, \tau_n\}$, where $\tau_l\in I_\pm$, $l=1,\dots,n$ and $\sum_{l=1}^n r(\tau_l)=k$. In other words, $I_n(k)$ denotes the set of paths that link the origin to $k\in \Z^d$ in $n$ steps via the jump function $r$. Let us write $\ffi_0=\sum_{\tau\in I_\pm}a_\tau |\tau\ket$.
\begin{lem}\label{wkn}
\bea
W_k(n)&=&\sum_{{\tau_0,{\{\tau_1,\cdots, \tau_n\}\in I_n(k)} \atop {\tau_0', \{\tau_1',\cdots, \tau_n'\}\in I_n(k)}} \atop \mbox{\tiny s.t.}\ \tau_n=\tau_n' }\overline{a_{\tau_0'}}a_{\tau_0}  \bra\tau_0'|C_1^*(0)\ \tau_1'\ket\bra \tau_1|C_1(0)\tau_0\ket \times \\\nonumber 
&&\hspace{2cm}\times \prod_{s=2}^n \Big\bra \tau_{s-1}'\Big|C_s^*\Big(\sum^{s-1}_{j=1}r(\tau_j')\Big)\ \tau_s'\Big\ket\Big\bra \tau_s\Big |C_s\Big(\sum^{s-1}_{j=1}r(\tau_j)\Big) \ \tau_{s-1}\Big\ket .
\eea
\end{lem}

We approach the problem through the characteristic functions $\Phi_n$ of the probability distributions $\{W_\cdot(n)\}_{n\in\N^*}$ defined by the periodic function
\be\label{char}
\Phi_n(y)=\E_{W(n)}(e^{i yX_n})=\sum_{k\in \Z^d}W_k(n)e^{i yk}, \ \ \mbox{where }\ y\in [0,2\pi)^d.
\ee
To emphasize the dependence in the initial state, we will sometimes write $\Phi_n^{\ffi_0}$ and/or $W_k^{\ffi_0}(n)$.
All periodic functions will be viewed as  functions defined on the torus, {\it i.e.} $[0,2\pi)^d \simeq \T^d.$
The asymptotic properties of the quantum walk emerge from the analysis of the limit in an appropriate sense as $n\ra\infty$ of the characteristic function in the {\em diffusive scaling}
\be\label{diffscal}
\lim_{n\ra\infty}\Phi_n(y/\sqrt n)
\ee

\section{Correlated Markovian Random Framework}\label{rands}

We give here the hypotheses we make on the randomness of the model.  \medskip\\ 
 {\bf Assumption C}:\\
Let $\Omega=\{C_1, C_2, \cdots, C_F\}$ be a finite set of unitary coin matrices on $\C^{2d}$ and let $\omega\in \Omega^\N$  be a Markov chain  with stationary initial distribution $p$ and transition matrix $\P$ s.t. $\P(\eta,\zeta)=\text{Prob}(\omega(n+1)=\zeta|\omega(n)=\eta)$, for all $n\in\N$. Let $\sigma$ be a representation of $\Z^{d}$, $x \mapsto \sigma_{x}$, in terms of measure preserving maps $\sigma_{x}:\Omega \ra \Omega$ such that $p(\sigma_x\zeta)=p(\zeta)$ and $\P(\sigma_x\zeta,\sigma_x\eta)=\P(\zeta,\eta)$. 

\begin{rems}  
i) This is equivalent to saying that the paths of $\sigma_{x}(\omega(\cdot))$ have the same distribution as the paths of $\omega(\cdot)$, for all $x \in \Z^{d}$.\\
ii) Because $x\mapsto \sigma_x$ is a representation of $\Z^d$, $\sigma_x$ is a bijection over the finite set $\Omega$ for any $x\in \Z^d$ and $\sigma_0=\operatorname{Id}$. Moreover, the finite set of bijections $\{\sigma_x\}_{x\in\Z^d}$ must commute with one another.\\
iii) Let $\Gamma=\{x\in \Z^d \ \mbox{s.t.}\ \sigma_x=\operatorname{Id} \}$. Then $\sigma_x=\sigma_y$ is equivalent to $x-y\in \Gamma$. If $g\in \N^*$ denotes the cardinal of the group  $\{\sigma_x\}_{x\in\Z^d}$, then for any $j\in\{1,\cdots, d\}$, the vector ${(0, \cdots, 0, g, 0, \cdots, 0)^T}\in \Z^d$, where $g$ sits at the $j$'th slot, belongs to $\Gamma$. Hence the lattice $\Gamma$ is of dimension $d$.\\
iv) We choose $B_\Gamma\subset \Z^d$ such that $0\in B_\Gamma$ and $\sigma|_{B_\Gamma}$ is a bijection on the set of bijections of $\Omega$. For any $x\in\Z^d$, we have a unique decomposition $x=x_0+\eta$, with $x_0\in B_\Gamma$ and $\eta\in \Gamma$.

\end{rems}

\medskip
We consider the random evolution obtained from sequences of coin matrices defined on site $x\in \Z^d$ at time $n\geq 0$ by 
\be\label{racm}
C^\omega_n(x)=\sigma_{x}(\omega(n)).
\ee
\medskip

This means that while the coin matrices at different sites have all the same distribution as $C^\omega_n(0)=\omega(n)$, they can take different correlated values depending on $\sigma_x$. \\

It is more natural  in this setting to carry out the analysis in terms of density matrices. The set of density matrices, ${\cal DM}$, consists in trace one non negative operators on  $\C^{2d}\otimes l^{2}(\Z^d)$.  Any bounded operators on $\cH=\C^{2d}\otimes l^{2}(\Z^d)$  can be represented by its kernel as
\be\label{op}
\rho=\sum_{(x, y)\in \Z^{2d}}\rho(x,y)\otimes |x\ket\bra y|, \ \ \mbox{where }\ \rho(x,y)\in M_{2d}(\C).
\ee
A non-negative operator $\rho$ on $\cH$ is trace class iff 
\be
\sum_{x\in\Z^d}\|\rho(x,x)\| <\infty. 
\ee
We say that $\rho$ belongs to $l^2(\Z^{d}\times\Z^d ; M_{2d}(\C))$ when 
\be\label{n2}
\sum_{(x,y)\in\Z^d\times\Z^d}\|\rho(x,y)\|^2<\infty.
\ee
Note that (\ref{n2}) is equivalent to the Hilbert-Schmidt norm induced  by the scalar product on 
 $l^2(\Z^{d}\times\Z^d ; M_{2d}(\C))$
\be
\big\bra \eta ,\, \rho \big\ket=\tr (\eta^* \rho)=\sum_{(x,y)\in\Z^d\times\Z^d}\tr (\eta(x,y)^*\rho(x,y)),
\ee
where we used the same symbol  ``$\tr$"  for the trace in different spaces, which make $l^2(\Z^{d}\times\Z^d ; M_{2d}(\C))$ a Hilbert space. We also note that if $\rho$
is non-negative, this implies for any $x,y\in\Z^d$, (see \cite{J1} and Lemma 1.21 in \cite{z}) 
\be \rho(x,y)=\rho(y,x)^*, \ \ \ 
\rho(x,x)\geq 0,\ \ \ \mbox{and} \ \ \ 
 \|\rho(x,y)\|\leq \|\rho(x,x)\|^{1/2}\|\rho(y,y)\|^{1/2}.
\ee 
Thus  ${\cal DM}$ and the set of non-negative trace-class operators belong to $l^2(\Z^{d}\times\Z^d ; M_{2d}(\C))$.\\

If $\rho_0$ denotes the initial density matrix, its evolution at time $n$ under 
$U(n,0)$ defined by (\ref{defUn}) is given by
\be
\rho_n=U(n,0)\rho_0U^*(n,0).
\ee
The kernel of $\rho_n$ reads
\be
\rho_n(x,y)=\sum_{(k,k')\in\Z^{d}\times\Z^d} J_k^{x-k}(n)\rho_0(x-k,y-k'){J_{k'}^{y-k'}}^*(n),
\ee
and the expectation of the lattice observable $F=\I\otimes f$ is denoted by
\be
\bra F\ket_{\rho_0}(n)=\tr (\rho_n (\I\otimes f))=\sum_{x\in\Z^d}\tr( \rho_n(x,x)) f(x),
\ee
if it exists. Again, we can express $\bra F\ket_{\rho_0}(n)$ as the expectation of  a random variable on the lattice $\Z^d$:
\be
\bra F\ket_{\rho_0}(n)=\E_{W(n)}(f(X_n)), \ \ \ \mbox{with } \ \ \ \mbox{ Prob}(X_n=k)=W_k(n)=\tr( \rho_n(k,k)).
\ee
In case the evolution is random, the distribution $W^\omega(n)$ is random and the density matrix $\rho_n^\omega$ is random  as well. We consider thus the expectation with respect to the randomness, noted $\E$ of the quantum mechanical expectation of the lattice observable, i.e.
\be
\E(\bra F\ket_{\rho_0}^\omega(n))=\E_{W^\omega(n)}(f(X_n))\equiv 
\E_{w(n)}(f(X_n)),
\ee
where the distribution $w(n)$ on $\Z^d$ is given by
\be\label{wW}
\mbox{ Prob}(X_n=k)=w_k(n)=\E(W^\omega_k(n))=\E \tr( \rho_n^\omega(k,k)).
\ee
The corresponding characteristic function is defined by
\be\label{averex}
\Phi_n^{\rho_0}(y)=\E_{w(n)}(e^{iy X_n})=\sum_{x\in\Z^d} e^{iyx} \tr( \,\E\,(\rho_n^\omega(x,x))).
\ee

The following assumption gives the required  regularity properties on the lattice observable $F=\I\otimes f$ and the initial density matrix $\rho_0$ to legitimate the manipulations that follow.

\medskip
{\bf Assumption R:} \\
a) The lattice observable is such that, for any $\mu<\infty$, $\exists C_\mu<\infty$ such that
\be
|f(x+y)|\leq C_\mu |f(x)|, \ \ \forall \, (x,y)\in \Z^d\times\Z^d \ \ \mbox{with}\, \|y\|\leq\mu.
\ee
b) The kernel $\rho_0(x,y)$ is such that 
\bea
&&\sum_{(x,y)\in \Z^{d}\times\Z^d}\| \rho_0(x,y)\|<\infty \\
&&\sum_{x\in\Z^d}|f(x)|\|\rho_0(x,x)\|<\infty.
\eea
Lemma 2.11 of \cite{J1} applies here as well, with the same proof, to ensure that for any $n\in \N$, the kernel $\rho_n(x,y)$ satisfies Assumption {\bf R} if the kernel $\rho_0$ does. 
For more discussion about properties of the density matrices we refer to \cite{J1}. 
\medskip

\subsection{ Feynmann-Kac-Pillet formula}

We denote by $l^2(\Omega; M_{2d}(\C))$ the finite dimensional Hilbert space of  $M_{2d}(\C)$-valued functions defined on $\Omega$ with scalar product defined by 
\be
\bra \ffi \, , \psi \ket=\sum_{\eta\in \Omega} p(\eta) \tr (\ffi^*(\eta)\psi(\eta)),
\ee
where the measure $p$ on $\Omega$ is the stationary initial distribution. We denote by $|\tau\ket\bra\tau '|\in l^2(\Omega; M_{2d}(\C))$ the constant map which assigns $|\tau\ket\bra\tau '|$ to any $\eta\in \Omega$ and stress that the $\tau, \tau'$ element of a matrix $\rho\in M_{2d}(\C)$, $\tau, \tau' \in I_\pm$, can be expressed as
\be
(\rho)_{\tau, \tau'}=\tr ( |\tau'\ket\bra\tau|\, \rho )= \tr ( (|\tau\ket\bra\tau '|)^*\, \rho).
\ee

Consider now the extended Hilbert space $l^2(\Z^{d}\times\Z^d; l^2(\Omega; M_{2d}(\C)))\simeq l^2(\Z^{d}\times\Z^d\times \Omega;M_{2d}(\C))$. Any
$\rho\in l^2(\Z^{d}\times\Z^d\times \Omega;M_{2d}(\C))$ can be expressed as 
\be
\rho=(\rho(x,y;\eta))_{(x,y;\eta)\in \Z^d\times\Z^d\times\Omega}, \ \ \mbox{where} \ \ \
\rho(x,y;\eta)\in M_{2d}(\C)
\ee
satisfies
\be
\sum_{{\eta\in \Omega} \atop {(x,y)\in\Z^d\times\Z^d}} p(\eta)\tr (\rho(x,y;\eta)^*\rho(x,y;\eta))<\infty.
\ee
 \medskip
 
The following is a version of Feynman-Kac-Pillet formula  in the current setting.
Let $\rho_0\in l^2(\Z^{d}\times\Z^d;M_{2d}(\C))$ denote the initial density matrix, its evolution at time $n$ under the random evolution operator
$U(n,0)$ defined by (\ref{defUn}) and (\ref{defJ}) is given by
\be
\rho_n=U(n,0)\rho_0U^*(n,0).
\ee
Since $l^2(\Z^{d}\times\Z^d;M_{2d}(\C))\hookrightarrow l^2(\Z^{d}\times\Z^d\times \Omega;M_{2d}(\C))$, we can consider $\rho_0$ an element of $ l^2(\Z^{d}\otimes\Z^d\times \Omega;M_{2d}(\C))$, keeping the same notation. With the notation $\delta_x=|x\ket$ we have
\begin{prop}\label{feykac}
Let $\mathcal{K}$ $=l^2(\Z^{d}\times\Z^d\times \Omega; M_{2d}(\C))$ and assume {\bf C} holds. Then, if $\rho_0\in \mathcal{K}$, we have for any $n\in\N$, and any $\tau, \tau' \in I_\pm$,
\be
\E(\rho_n (x,y))_{\tau, \tau'} = \Bra {\delta_x}\otimes {\delta_y} \otimes |\tau\ket\bra\tau '| \, , M^n \rho_0  \Ket_{\mathcal{K}}, \label {E(Rho)}
\ee
where  the single step operator $M:\cK\ra\cK$ is given by
\be\label{defM}
(M\rho)(x,y;\eta)=\sum_{\tau,\tau'\in I_\pm\atop{\zeta\in\Omega}} \Q(\eta,\zeta)P_\tau (\sigma_{x-r(\tau)}\eta) \rho(x-r(\tau),y-r(\tau'),\zeta)(\sigma_{y-r(\tau')}\eta)^* P_{\tau'},
\ee
where  $\rho\in l^2(\Z^{d}\times\Z^d\times \Omega; M_{2d}(\C))$ and  $\Q(\eta,\zeta)=\text{\em Prob}(\omega(0)=\eta|\omega(1)=\zeta)$.
\end{prop}
\begin{rems} i) Using that the initial distribution is stationary, it is easy to see that 
\be\label{QandP}
 \Q(\zeta,\eta)=\dfrac{p(\eta)}{p(\zeta)}\P(\eta,\zeta).
\ee
ii) In view of (\ref{wW}), the averaged distribution $w(n)$ reads
\be\label{psix}
w_x(n)=\sum_{\tau\in I_\pm}\E(\rho_n (x,x))_{\tau, \tau}=\Bra \Psi_x, \, M^n\rho_0\Ket
\ \ \ \mbox{where} \ \ \  
\Psi_x={\delta_x}\otimes {\delta_x} \otimes \operatorname{Id}.
\ee
iii) The adjoint of $M$, $M^*$, acts as follows
\be\label{defM*}
(M^*\rho)(x,y;\eta)=\sum_{\tau,\tau'\in I_\pm\atop{\zeta\in\Omega}} \P(\eta,\zeta) (\sigma_{x}\zeta)^*P_\tau \rho(x+r(\tau),y+r(\tau'),\zeta)P_{\tau'}(\sigma_{y}\zeta).
\ee
iv) If $\{\rho(x,y;\eta)\}_{x,y\in\Z^d}$ is self-adjoint, the same is true for $\{(M\rho)(x,y;\eta)\}_{x,y\in\Z^d}$. Such initial conditions $\rho$ yield real valued quantities $w_x(n)=\Bra \Psi_x, \, M^n\rho_0\Ket$.
\end{rems}
\proof First note that 
\be
\big\langle\delta_{x}\otimes \delta_{y} \otimes |\tau\ket\bra\tau '| \, , M^n \rho_0 \big\rangle_{\mathcal{K}}=\sum_{\zeta\in\Omega}p(\zeta) ((M^n \rho_0)(x,y;\zeta))_{\tau, \tau'}.
\ee
 Let $t_i=\sum_{s=i}^n r(\tau_s)$ and $t_i'=\sum_{s=i}^n r(\tau_s')$. Using the definition of $M$, we see that
\bea
&& (M^n \rho_0)(x,y;\zeta)=\sum_{{{\{\tau_1,\cdots, \tau_n\}\in I_n} \atop { \{\tau_1',\cdots, \tau_n'\}\in I_n}} \atop {\{\eta_1,\eta_2,\cdot,\eta_n}\}\in \Omega^n } \Q(\zeta,\eta_n)\Q(\eta_n,\eta_{n-1})\cdots \Q(\eta_2,\eta_1)\times \nonumber 
 \\ \nonumber 
 && \hspace{0cm}P_{\tau_n}(\sigma_{x-t_n}\zeta)P_{\tau_{n-1}}(\sigma_{x-t_{n-1}}\eta_n)\cdots 
 P_{\tau_1}(\sigma_{x-t_1}\eta_2)\rho_0(x-t_1,y-t'_1) \times\\ \nonumber 
 &&\hspace{5cm}(\sigma_{y-t_1'}\eta_2)^*P_{\tau'_1}\cdots (\sigma_{y-t_{n-1}'}\eta_n)^*P_{\tau'_{n-1}}(\sigma_{y-t_n'}\zeta)^*P_{\tau'_n}.\nonumber
\eea
Since the initial distribution $p$ is stationary, a straightforward computation shows that 
\be
p(\zeta)\Q(\zeta,\eta_n)\Q(\eta_n,\eta_{n-1})\cdots \Q(\eta_2,\eta_1)=p(\eta_1)\P(\eta_1,\eta_2)\cdots \P(\eta_{n-1},\eta_n)\P(\eta_n,\zeta).
\ee
Therefore,
\bea\label{matrixn}
&& \big\langle\delta_{x}\otimes \delta_{y} \otimes | \tau \ket \bra \tau' | \, , M^n \rho_0 \big\rangle_{\mathcal{K}}=\sum_{{{\{\tau_1,\cdots, \tau_n\}\in I_n} \atop { \{\tau_1',\cdots, \tau_n'\}\in I_n}} \atop {\{\eta_1,\eta_2,\cdot,\eta_n,\zeta}\}\in \Omega^{n+1} } p(\eta_1)\P(\eta_1,\eta_2)\cdots\P(\eta_{n-1},\eta_n)\P(\eta_n,\zeta)\times \nonumber
\\ \nonumber
&& \bra \tau | P_{\tau_n}(\sigma_{x-t_n}\zeta)P_{\tau_{n-1}}(\sigma_{x- t_{n-1}}\eta_n)\cdots 
P_{\tau_1}(\sigma_{x-t_1}\eta_2)\rho_0(x-t_1,y-t'_1)\times 
\\
&&\hspace{4.5cm}(\sigma_{y-t'_1}\eta_2)^* P_{\tau'_1}\cdots (\sigma_{y-t_{n-1}'}\eta_n)^*P_{\tau'_{n-1}}(\sigma_{y-t_n'}\zeta)^*P_{\tau'_n}\ \tau' \ket.
\eea

On the other hand,
\bea
&&\E(\rho_n (x,y))=\sum_{{{\{\tau_1,\cdots, \tau_n\}\in I_n} \atop { \{\tau_1',\cdots, \tau_n'\}\in I_n}}\atop { \{\eta_1,\eta_2,\cdot,\eta_n\}\in \Omega^n\atop{ \{\eta'_1,\eta'_2,\cdot,\eta'_n\}\in \Omega^n}} } \text{Prob}\left( \sigma_{x-t_i}\omega(i)=\eta_i, \sigma_{y-t'_i}\omega(i)=\eta'_i\text{ for all } i\in\{1,\cdots,n\}\right)\nonumber
\\ &&\hspace{2cm}\times\ P_{\tau_n}\eta_n P_{\tau_{n-1}}\eta_{n-1}\cdots P_{\tau_1}\eta_1\rho_0\left(x-t_1,y-t'_1\right)\eta'^*_1 P_{\tau'_1}\cdots P_{\tau'_{n-1}}\eta'^*_n P_{\tau'_{n}}.
\eea
However, it is easy to see that
\bea
 \text{Prob}\left( \sigma_{x-t_i}\omega(i)
=\eta_i, \sigma_{y-t'_i}\omega(i)=\eta'_i\text{ for all } i\in\{1,\cdots,n\}\right)\nonumber
\\=\begin{cases}
\text{Prob}\left(\omega(i)=\sigma^{-1}_{x-t_i}\eta_i\text{ for all } i\in\{1,\cdots,n\}\right)&\text{if   } \sigma^{-1}_{x-t_i}\eta_i=\sigma^{-1}_{y-t'_i}\eta'_i
\\0& \text{otherwise.} 
\end{cases}
\eea
Now by letting $\alpha_i=\sigma^{-1}_{x-t_i}\eta_i$, and using that $\omega$ is  a Markov chain  on $\Omega$ we get

\bea\label{expectn}
&&\E(\rho_n (x,y))=\sum_{{{\{\tau_1,\cdots, \tau_n\}\in I_n} \atop { \{\tau_1',\cdots, \tau_n'\}\in I_n}}\atop  \{\alpha_0,\alpha_1,\cdot,\alpha_n\}\in \Omega^{n+1}}p(\alpha_0)\P(\alpha_0, \alpha_1)\P(\alpha_1, t\alpha_2)\cdots \P(\alpha_{n-1},\alpha_n) \times 
\\ \nonumber
&&\hspace{1cm}P_{\tau_n}(\sigma_{x-t_n}\alpha_n)P_{\tau_{n-1}}\cdots P_{\tau_1}(\sigma_{x-t_1}\alpha_1)
\rho_0\left(x-t_1,y-t'_1)\right)\times 
\\ \nonumber
&&\hspace{6cm}(\sigma_{y-t'_1}\alpha_1)^*P_{\tau'_1}\cdots P_{\tau'_{n-1}}(\sigma_{y-t'_n}\alpha_1)^*P_{\tau'_n}.
\eea

Comparing (\ref{matrixn}) and (\ref{expectn}) completes the proof.\ep

\medskip
\subsection{Spectral Analysis of $M$}\label{sp}
Using Feynman-Kac-Pillet formula, studying the time evolution of our systems relies on the spectral analysis of the ``single-step"  operator $M$ defined in (\ref{defM}).
 In order to do that we first take a closer look at the underlying symmetries of the systems. The operator $M$ commutes with a group $\cG$ of unitary operators generated by translations:
\begin{enumerate}
\item Simultaneous translation of position and disorder by an arbitrary element $\xi$ of $\Z^{d}$:
$$ S_{\xi}\,\rho(x,y; \omega)=\rho(x-\xi,y-\xi; \sigma_{\xi}\omega),$$
\item For $\eta \in\Gamma\subset \Z^d$ such that $\sigma_\eta=\operatorname{Id}$ ,  $M$ commutes with translation of the first position coordinate by $\eta$:
$$ S^{(1)}_{\eta}\rho(x,y; \omega)=\rho(x-\eta, y; \omega).$$
\end{enumerate}
 Note that $S_{\xi}S^{(1)}_{\eta}=S^{(1)}_{\eta}S_{\xi}$, so the group of symmetries $\mathcal{G}$ is isomorphic to $\Z^{d}\times \Z^{d}$.   We have chosen to use translation of the first position in the definition of $S^{(1)}$; however, since $\sigma_{\eta}=\operatorname{Id}$, we have $S_{\eta}S^{(1)}_{-\eta}\rho(x,y;\omega)=\rho(x,y-\eta; \omega)$.
\begin{rem}\label{x0}
For any $\eta\in \Gamma=\{\xi\in\Z^d: \sigma_\xi=\operatorname{Id}\}$, and any  $x\in\Z^d$, we have $\sigma_{x+\eta}=\sigma_x$. Moreover, for any $x\in \Z^d$, there exists a unique $x_0\in B_\Gamma$ such that $\sigma_x=\sigma_{x_0}$.

\end{rem}

\medskip
 
In order to  take these symmetries into account in the spectral analysis of $M$, we define  a generalized Fourier transform similar to \cite{hks}. We shall use the following notations 
$$L^2(X;M_{2d}(\C))=\{f:X\to M_{2d}(\C): \|f\|^2=\mbox{Tr}(f^*f)=\int_{X} dm(x)\mbox{Tr}(f^*(x)f(x))<\infty \},$$
where $m$ is a locally finite positive measure on $X$.

Also, we introduce $\Gamma^*=\{p^*\in \R^d\ | \ \ p^*\gamma\in \Z, \ \forall \gamma\in \Gamma\}$. If $\{\gamma_j\}_{j=1}^d$ is a basis of $\Gamma$, let $\{p_j^*\}_{j=1}^d$ be the basis of $\Gamma^*$ defined by $p_j^*\gamma_i=\delta_{i,j}$. We have that $\Z^d\subset \Gamma^*$. We note $\T_{\Gamma}^d=\{p=\sum_{j=1}^d p_j p_j^* \ | \ p_j\in [0,2\pi), j=1,\dots, d\}$. With the linear map $P: \R^d\ra\R^d$ defined  by its action on the vectors of the canonical basis as $Pe_j=p^*_j$, $j=1, \dots, d$, we have $\T_{\Gamma}^d=P\T^d$ where $\T^d$ denotes the $d$-dimensional torus. In particular, for any $f$ defined on $\T_\Gamma^d$
\be
\int_{{\T_{\Gamma}^d}}f(p)dp =\int_{\T_d}f(Pt)|P|dt, \ \ \mbox{and} \ \ \ \mbox{Vol}(T_\Gamma^d)=(2\pi)^d|P|,
\ee
where $|\cdot |$ denotes the Jacobian determinant here. We denote the normalized measure on $\T_\Gamma^d$ by $d\tilde p=dp/((2\pi)^d|P|)$.

\medskip

We are ready to introduce the map ${\cal F}: l^2(\Z^{d} \times \Z^{d} \times \Omega;M_{2d}(\C)) \to L^2(B_\Gamma \times \T^d\times \T_\Gamma^d\times \Omega;M_{2d}(\C) )$, defined by
\begin{equation}
(\cF \Psi)(x,k,p;\zeta):=\widehat{ \Psi}(x,k,p;\zeta)  = \sum_{{\xi \in \Z^{d}}\atop{\eta\in\Gamma} } e^{i p \cdot (x-\eta) -i k\cdot \xi}  \Psi (x -\xi -\eta , -\xi  , \sigma_{\xi}\zeta).
\end{equation} 
Since we can add to $x$ any vector of $\Gamma$ without changing the RHS, $\widehat \Psi$ actually depends on $x_0\in B_\Gamma$ defined according to remark  \ref{x0}. 
One checks that this generalized Fourier transform is a unitary operator with inverse 
\be
(\cF^{-1} \chi)(x,y;\zeta)=\int_{\T^d\times\T_\Gamma^d}e^{-iyk}e^{-ip(x-y)}\chi(x-y,k,p;\sigma_y\zeta)\tilde dk \tilde dp,
\ee
where $\tilde dk \tilde dp$ is the normalized measure on $\T^d\times \T_\Gamma^d$.
\begin{rems} 
i) If $(\cF \Psi)(x,k,p;\zeta)=\widehat{ \Psi}(x_0,k,p;\zeta)$, it satisfies for any $p^*\in\Gamma^*$, any $\eta\in\Gamma$ and any $k^*\in \Z^d$
\be
\widehat{ \Psi}(x_0+\eta,k+2\pi k^*,p+2\pi p^*;\zeta)=e^{i2\pi p^*x_0}\widehat{ \Psi}(x_0,k,p;\zeta).
\ee \label{remsym}
ii) The operator $\{\psi(x,y,\zeta)\}_{x,y\in\Z^d}$ is self-adjoint, i.e. $\psi(x,y,\zeta)^*=\psi(y,x,\zeta)$ if and only if  $\widehat{ \Psi}(x_0,k,p;\zeta)=\widehat{ \Psi}((-x)_0,-k,p-k;\sigma_{x_0}\zeta)^*$.
\end{rems}

Because of the symmetries of $M$, its expression $\cF M \cF^{-1}$ in Fourier space admits a fiber decomposition of the form
\be
\cF M \cF^{-1}=\int^\oplus_{\T^d\times\T_\Gamma^d} \widehat M(k,p) \, dk dp,
\ee
where $\widehat M(k,p)$ is an operator on $l^2(B_\Gamma\times\Omega; M_{2d}(\C))$ which becomes a multiplication operator in the variables $(k,p)\in \T^d\times\T_\Gamma^d$ which we compute. The following expression holds for the $(k,p)$ dependent  "single-step" operator $\widehat M(k,p)$ on $ L^2(B_\Gamma\times\T^d\times\T_\Gamma^d\times\Omega; M_{2d}(\C) )$:
\bea\label{fourierM}
&&(\widehat M(k,p)\Psi)(x,k,p;\eta)=\sum_{\tau,\tau'\in I_\pm\atop{\zeta\in\Omega}} \Q(\eta,\zeta)e^{ikr(\tau')} e^{ip(r(\tau)-r(\tau'))}\times 
\\ \nonumber
&&\hspace{4cm}P_\tau (\sigma_{x-r(\tau)}\eta) \Psi\left(x-r(\tau)+r(\tau'),k,p,\sigma_{-r(\tau')}\zeta\right)(\sigma_{-r(\tau')}\eta)^*P_{\tau'}.
\eea
\begin{rem}
The action of the adjoint of $\widehat M(k,p)$, denoted by $\widehat M(k,p)^*$, reads
\bea
&&(\widehat M(k,p)^*\Psi)(x,k,p;\eta)=\sum_{\tau,\tau'\in I_\pm\atop{\zeta\in\Omega}} \P(\eta,\zeta)e^{-ikr(\tau')} e^{-ip(r(\tau)-r(\tau'))}\times 
\\ \nonumber
&&\hspace{4cm}(\sigma_{x}\zeta)^*P_\tau  \Psi\left(x+r(\tau)-r(\tau'),k,p,\sigma_{r(\tau')}\zeta\right)P_{\tau'} \zeta.
\eea
\end{rem}
\medskip

Let us now consider the operator $\widehat M(k, p)$ for $(k,p)\in\T^d\times\T_\Gamma^d$ fixed as an operator on $l^2(B_\Gamma\times\Omega; M_{2d}(\C))$. As $B_\Gamma$ and $\Omega$ are finite, $\widehat M(k, p)$ can be represented by a square matrix of dimension $4d|B_\Gamma| |\Omega| $ depending parametrically on $(k,p)$. Moreover, the map $(k,p)\mapsto \widehat M(k, p)$ is analytic on  $\C^d\times\C^d$.
We denote the norm on $l^2(B_\Gamma\times\Omega; M_{2d}(\C))$ by $\|\cdot \|_{l^2}$.

\begin{prop}
Let ${\em \spr}$ denote the spectral radius, then for all $(k,p)\in\T^d\times\T_\Gamma^d$ 
\be\label{bddspec}
{\em \spr}(\widehat M(k,p))\leq \|\widehat M(k,p)\|_{l^2}\leq 1,
\ee
On the other hand for $k=0$ and all $p\in\T_\Gamma^d$
\be\label{spn}
{\em \spr}(\widehat M(0,p))= \|\widehat M(0,p)\|_{l^2}= 1,
\ee
\end{prop}
\begin{rem}
It follows that ${\em \spr} (\widehat M)=\|\widehat M\|=1$, where $\widehat M$ is viewed as an operator on $ L^2(B_\Gamma \times \T^d\times \T_\Gamma^d\times \Omega;M_{2d}(\C) )$.
\end{rem}

\proof First note that $\widehat M(k,p)$ can be written as $\widehat M(k,p)=\Sigma S\widetilde{Q}$, where
\bea \label{decm}\nonumber
(\widetilde{Q}\Psi)(x,k,p;\eta)&=&\sum_{\zeta\in\Omega} \Q(\eta,\zeta)\Psi(x,k,p;\zeta)
\\
(S\Psi)(x,k,p;\eta)&=&(\sigma_x\eta) \Psi(x,k,p;\eta) (\sigma_0\eta)^*
\\  \nonumber
(\Sigma\Psi)(x,k,p;\zeta)&=&\sum_{\tau,\tau'\in I^2_\pm} e^{ikr(\tau')} e^{ip(r(\tau)-r(\tau'))}P_\tau\Psi(x-r(\tau)+r(\tau'),k,p;\sigma_{-r(\tau')}\zeta)P_{\tau'}.
\eea
We fix $(k,p)$ and consider these operators on $l^2(B_\Gamma\times\Omega; M_{2d}(\C))$. 
Now $\widetilde{Q}=\operatorname{Id}\otimes Q$ where $Q:l^2(\Omega; \C)\to l^2(\Omega; \C)$  given by
$Qf(\eta)=\sum_{\zeta\in\Omega} \Q(\eta,\zeta)f(\zeta)$ and $\operatorname{Id}$ means the identity on $l^2(B_\Gamma; M_{2d}(\C))$. An easy calculation using Jensen's inequality shows that for all $f\in l^2(\Omega; \C)$
\bea
\|Qf\|^2=\sum_{\eta\in\Omega}p(\eta)\left|\sum_{\zeta\in\Omega}\Q(\eta,\zeta) f(\zeta)\right|^2
\leq \sum_{\eta\in\Omega}p(\eta)\sum_{\zeta\in\Omega}\Q(\eta,\zeta) |f(\zeta)|^2
\\=\sum_{\zeta\in\Omega}p(\zeta) |f(\zeta)|^2=\|f\|^2.
\eea
Therefore, we have that $\|\widetilde{Q}\|_{l^2}\leq 1$.

On the other hand, for all $\Psi \in l^2(B_\Gamma \times  \Omega; M_{2d}(\C) )$
\bea\nonumber
\|S\Psi\|_{l^2}^2&=&\sum_{x\in B_\Gamma \atop{\zeta\in \Omega}}p(\zeta)\mbox{Tr}\left[ \zeta\Psi^*(x;\eta) (\sigma_x\zeta)^*(\sigma_x\zeta)\Psi(x;\eta) \zeta^*\right]
\\&=&\sum_{x\in B_\Gamma \atop{\zeta\in \Omega}}p(\zeta)\mbox{Tr}\left[ \Psi^*(x;\eta) \Psi(x;\eta)\right]=\|\Psi\|_{l^2}^2.
\eea
Where we used the cyclicity of the trace and that elements of $\Omega$ are unitary matrices.
Finally to see that for any $(k,p)\in\T^d\times \T_\Gamma^d$, $\|\Sigma\|=1$, we notice that
\bea
 &&\mbox{Tr}\left[ (\Sigma\Psi)^*(x;\zeta)(\Sigma\Psi)(x;\zeta)\right]=
\\ \nonumber
&&\sum_{\tau,\alpha\in I^2_\pm} \langle\alpha| \Psi^*(x-r(\tau)+r(\alpha);\sigma_{-r(\alpha)}\zeta)\tau\rangle  \langle\tau| \Psi(x-r(\tau)+r(\alpha);\sigma_{-r(\alpha)}\zeta)\alpha\rangle 
\eea
Now  for fixed $\alpha, \tau$ let $y=x-r(\tau)+r(\alpha)$ and $\eta=\sigma_{-r(\alpha)}\zeta$. Using that $\sigma_x$ are measure preserving transformations on $\Omega$, we have
\bea
\|\Sigma\Psi\|_{l^2}^2=\sum_{\tau,\alpha\in I^2_\pm} \sum_{y\in B_\Gamma \atop{\eta\in \Omega}}p(\eta)\langle\alpha| \Psi^*(y;\eta)\tau\rangle \langle\tau| \Psi(y;\eta)\alpha\rangle 
\\= \sum_{y\in B_\Gamma \atop{\eta\in \Omega}}p(\eta)\mbox{Tr}\left[ \Psi^*(y;\eta) \Psi(y;\eta)\right]=\|\Psi\|^2.\label{normSigma}
\eea
Putting the estimates on the norms of $\widetilde{Q}$, $S$ and $\Sigma$ together we get the required bound on the norm of $\widehat M(k,p)$ for all $(k,p)\in\T^d\times\T_\Gamma^d$.

Now  consider $\widehat\Psi_1(x;\zeta)=\delta_0\otimes \operatorname{Id}$, where $\operatorname{Id}\in l^2(\Omega;M_{2d}(\C))$ takes the constant value  $\operatorname{Id}$. We compute
\bea\nonumber
(\widehat M(k,p)\widehat\Psi_1)(x;\eta)&=&\sum_{\tau,\tau'\in I_\pm\atop{\zeta\in\Omega}}e^{ikr(\tau')} e^{ip(r(\tau)-r(\tau'))} \Q(\eta,\zeta) \times \\ \nonumber
&&\hspace{3cm}P_\tau (\sigma_{x-r(\tau)}\eta) (\sigma_{-r(\tau')}\eta)^*P_{\tau'}\delta_0(x-r(\tau)+r(\tau'))\\ \nonumber
&=&\sum_{\tau,\tau'\in I_\pm}e^{ikr(\tau')} e^{ip(r(\tau)-r(\tau'))} P_\tau \delta_{\tau,\tau'}\delta_0(x-r(\tau)+r(\tau'))
\\ 
&=&\sum_{\tau\in I_\pm} e^{ikr(\tau)}P_\tau \delta_0(x).
\eea
 where we use $\sum_{\zeta\in\Omega}\Q(\eta,\zeta)=1$. From this it is clear that $\widehat\Psi_1$  is an eigenvector of $\widehat M(0,p)$ with an eigenvalue $1$ for all $p\in\T_\Gamma^d$. Therefore we have that 
\be
{ \spr}(\widehat M(0,p))= \|\widehat M(0,p)\|_{l^2}= 1.
\ee
\ep
\begin{rem}
A  similar computations shows that
\be\label{msp1}
\widehat M(0,p)^*\widehat\Psi_1=\widehat\Psi_1.
\ee
If  $\widehat\Psi_1$ is considered a vector of $L^2(B_\Gamma\times\T^d\times\T_\Gamma^d\times\Omega;M_{2d}(\C))$ it corresponds to 
\be
\Psi_1(x,y;\eta)=\cF^{-1}\widehat\Psi_1(x,y;\eta)=\delta_0(x)\otimes\delta_0(y)\otimes \operatorname{Id}\simeq\operatorname{Id}\otimes |0\ket\bra 0|.
\ee
Also, with the definition (\ref{psix})
\be\label{fpsix}
\cF \Psi_{\tilde x}(x,k,p,\eta)=e^{ik\tilde x}\delta_0(x)\otimes\operatorname{Id}=e^{ik\tilde x}\widehat \Psi_1(x).
\ee
\end{rem}

\medskip

At this point we note that the characteristic function $\Phi_n^{\rho_0}(y)$ of the distribution $w(n)$ satisfies, see (\ref{psix}) and (\ref{fpsix})
\bea
\Phi_n^{\rho_0}(y)&=&\sum_{x\in\Z^d}e^{iyx}\Bra \Psi_x, \, M^n\rho_0\Ket =\sum_{x\in\Z^d}e^{iyx}\Bra \widehat \Psi_x, \, \widehat M(\cdot , \cdot )^n\widehat \rho_0\Ket_{\cK} \\ \nonumber
&=&\sum_{x\in\Z^d}e^{iyx}\int_{\T^d}e^{-ikx}\Bra \widehat \Psi_1, \, \widehat M(k,\cdot)^n
\widehat \rho_0(k)\Ket_{L^2(B_\Gamma\times \T_\Gamma^d\times\Omega;M_{2d}(\C))}\widetilde dk.
\eea
In other words, slightly abusing notations,
\bea\label{iow}
\Phi_n^{\rho_0}(y)&=&\Bra \widehat \Psi_1, \, \widehat M(y,\cdot)^n
\widehat \rho_0(y)\Ket_{L^2(B_\Gamma\times \T^d\times\Omega;M_{2d}(\C))}
\nonumber\\
&=&
\Bra  ({\widehat {M}(y,\cdot)^*})^n\delta_0\otimes\operatorname{Id}, \,
\widehat \rho_0(y)\Ket_{L^2(B_\Gamma\times \T_\Gamma^d\times\Omega;M_{2d}(\C))}
\nonumber\\
&=&\int_{\T_\Gamma^d}\Bra  (\delta_0\otimes\operatorname{Id}, \, {\widehat {M}(y,p)})^n
\widehat \rho_0(y,p)\Ket_{l^2(B_\Gamma\times\Omega;M_{2d}(\C))}\widetilde dp
\nonumber\\
&=&\sum_{\eta\in \Omega}p(\eta)\int_{\T_\Gamma^d}
\tr (\widehat M(y,p)^n \widehat \rho_0)(0, y, p, \eta)
\widetilde dp
\eea
where
\be\label{ideta}
\widehat \rho_0(x,k,p,\zeta)= \sum_{{\xi\in\Z^d}\atop{\eta\in \Gamma}}e^{ip(x-\eta)-ik\xi}
\rho_0(x-\xi-\eta, -\xi)\equiv \widehat \rho_0(x,k,p)
\ee
is independent of $\zeta$.
\begin{rem} If
 \be\label{simple} 
\rho_0=|\ffi_0\ket\bra\ffi_0|\otimes |0\ket\bra0|\simeq\delta_0\otimes\delta_0\otimes|\ffi_0\ket\bra\ffi_0|,
\ \ \ \mbox{where $\ffi_0\in \C^d$ is normalized,}
\ee 
then
\be\label{rz}
\widehat \rho_0(x,k,p,\eta)= \delta_0(x)\otimes |\ffi_0\ket\bra\ffi_0|:=R_0 (x)\ \ \ \mbox{is independent of $(k,p,\eta)$ }
\ee
and
\be
\Phi_n^{\ffi_0}(y)=\sum_{\eta\in \Omega}p(\eta)\int_{\T_\Gamma^d}
{\em \tr} (\widehat M(y,p)^n |\ffi_0 \ket\bra\ffi_0 |) (0, y, p, \eta)
\widetilde dp.
\ee
\end{rem}

\medskip
Hence, in the diffusive scaling, we need to control the large $n$ behavior of the vectors $\widehat M^*(y/\sqrt n, p)^n \delta_0\otimes \operatorname{Id}$
 and $\widehat \rho_0(y/\sqrt n)$ in  $L^2(B_\Gamma\times \T_\Gamma^d\times\Omega;M_{2d}(\C))$. This can be done, following the arguments of \cite{J1}, under some spectral hypothesis. We shall discuss the validity of this hypothesis for specific cases later on and proceed by showing that it is sufficient to prove the diffusive character of the (averaged) dynamics, arguing as in \cite{J1}. We shall refrain from spelling out all details, referring the reader to the above mentioned paper.

\medskip

We work under the following spectral hypothesis on the matrix  $\widehat M(0,p)$ 
on $l^2(B_\Gamma\times\Omega;M_{2d}(\C))$. Let $D(z,r)$ be the open disc of radius $r>0$ centered at $z\in \C$. \\

{\bf Assumption S}: For all $p\in\T_\Gamma^d$, we have
\be
\sigma(\widehat M(0,p))\cap {\partial D(0,1)}=\{1\} \ \ \ \mbox{and this eigenvalue is simple.}
\ee
\begin{rem}\label{mresi}
Actually, because of (\ref{iow}), it is enough to assume that
$\widehat M(0,p))|_{\cI}$ satisfies assumption {\bf S}, where ${\cI}$ is the $\widehat M(k,p)^*$-cyclic subspace generated by $\delta_0\otimes\operatorname{Id}$, for all $(k,p)\in \T^d\times\T_\Gamma^d$.
\end{rem}

By analytic perturbation theory, there exist $\delta>0$, $\nu(\delta)>0$ and $\kappa(\delta)>0$ such that for all $(k,p)\in\cB^d_\kappa\times\cT^d_\nu$, where $\cB^d_\kappa=\{y\in \C^d \ | \ \|y\|<\kappa \}$ and $\cT^d_\nu=\{y=y_1+iy_2, | y_1\in \T_\Gamma^d, y_2\in\R^d\ \ \mbox{with}\ \ \|y_2\|< \nu \}$ the following holds:
\bea
&&\sigma(\widehat M(k,p))\cap D(1,\delta)=\{\lambda_1(k,p)\} \nonumber\\
&&\sigma(\widehat M(k,p))\setminus\{\lambda_1(k,p)\}\subset D(0,1-\delta),
\eea
and the eigenvalue $\lambda_1(k,p)$  is  simple. For such values of the parameters $(k,p)$ we have the corresponding spectral decomposition
\be\label{specdec}
\widehat M(k,p)=\lambda_1(k,p)P_1(k,p)+\widehat M_{\bar P_1}({k,p})
\ee
where $\widehat M_{\bar P_1}({k,p})={\bar P_1}({k,p})\widehat M(k,p){\bar P_1}({k,p})$ and ${\bar P_1}({k,p})=\I -P_1(k,p)$.

The simple eigenvalue $\lambda_1(k,p)$, the corresponding spectral projector $P_1(k,p)$ and the restriction $\widehat M_{\bar P_1}({k,p})$ are all analytic on $\cB^d_\nu\times\cT^d_\nu$ and $\spr (\widehat M_{\bar P_1}({k,p}))<1-\delta$. Moreover, for any $p\in\T_\Gamma^d$
\be\label{conspecdec}
\lim_{k\ra 0}\lambda_1(k,p)=1, \ \ \ \mbox{and} \ \ \ \lim_{k\ra 0}P_1(k,p)={|\widehat \Psi_0\ket\bra\widehat \Psi_0|}\equiv \Pi,
\ee
where $\widehat \Psi_0=\widehat \Psi_1/\|\widehat \Psi_1\|=\frac{1}{\sqrt{2d}}\delta_0\otimes\operatorname{Id}$, see (\ref{msp1}).
\medskip

Taking into account the fact that  $w_x(n)$ is real valued for any  selfadjoint and trace class $\rho_0$, we have
\be
\Phi_n^{\rho_0}(k)=\overline{\Phi_n^{\rho_0}(-k)}, \ \ \mbox{for all} \ \ k\in \T^d. 
\ee
This yields a symmetry of $\lambda_1$:
\begin{lem}\label{rvl}
For all $k \in \cB_\kappa^d$, and all $p\in\T_\Gamma^d$, the following identity holds
\be
\lambda_1(k,p)=\overline{\lambda_1(-\overline{k},p-\overline{k})}.
\ee
\end{lem}
\proof  It follows from (\ref{iow}) that for any $k\in\R^d$ and any self adjoint trace class $\rho_0$ 
\bea\label{symsym}
\Phi_n^{\rho_0}(k)&=&\int_{\T_\Gamma^d}\Bra  \delta_0\otimes\operatorname{Id}, \, {\widehat {M}(k,p)}^n
\widehat \rho_0(k,p)\Ket_{l^2(B_\Gamma\times\Omega;M_{2d}(\C))}\widetilde dp\nonumber\\
&=&\int_{\T_\Gamma^d}\overline{\Bra  \delta_0\otimes\operatorname{Id}, \, {\widehat {M}(-k,p)}^n
\widehat \rho_0(-k,p)\Ket}_{l^2(B_\Gamma\times\Omega;M_{2d}(\C))}\widetilde dp
\eea
The first step consists in showing the pointwise identity of the smooth scalar products in ${l^2(B_\Gamma\times\Omega;M_{2d}(\C))}$ for $\widehat \rho_0=\delta_0(x)\otimes |\ffi_0\ket\bra\ffi_0|=R_0(x):$
\be\label{point}
\Bra  \delta_0\otimes\operatorname{Id}, \, {\widehat {M}(k,p)}^n
 R_0\Ket-
\overline{\Bra  \delta_0\otimes\operatorname{Id}, \, {\widehat {M}(-k,p-k)}^n
R_0\Ket}=0.
\ee 
Identity (\ref{symsym}) holds for any self-adjoint $\rho_0$, thus in particular for 
$\rho_0(x_0,k,p)=b(k,p)R_0(x)$ where $b$ belongs to the vector space of periodic functions satisfying
\be
b:\T^d\times\T_\Gamma^d\ra \C, \ \ \ \mbox{such that }\ \ \ \overline{b(k,p)}=b(-k,p-k),
\ee
see Remarks \ref{remsym}. Therefore, we get for any such $b$
\bea\label{difbar}
0&=&\int_{\T_\Gamma^d}\Bra  \delta_0\otimes\operatorname{Id}, \, {\widehat {M}(k,p)}^n
 R_0\Ket b(k,p)-\overline{\Bra  \delta_0\otimes\operatorname{Id}, \, {\widehat {M}(-k,p)}^n
 R_0\Ket}b(k,p+k)\widetilde dp\nonumber\\
&=&\int_{\T_\Gamma^d}\Big (\Bra  \delta_0\otimes\operatorname{Id}, \, {\widehat {M}(k,p)}^n
 R_0\Ket-\overline{\Bra  \delta_0\otimes\operatorname{Id}, \, {\widehat {M}(-k,p-k)}^n
 R_0\Ket}\Big ) b(k,p)\widetilde dp.
\eea 
An example of smooth function $b$ satisfying our requirements is $b_1(k,p)=f_1(k)g_1(2p-k)$ with
\be
 g_1: \R^d\ra \R, \ 2\pi\Gamma^*-\mbox{periodic},\ \  \mbox{and} \ \ f_1: \R^d\ra \R, \ 2\pi\Z^d-\mbox{periodic and even}. 
\ee
Note that $\Z^d\subset \Gamma^*$ ensures $\Z^d$ periodicity of $b_1$ in $k$ and that $b_1(k,p+2\pi\gamma^*/2)=b_1(k,p)$, for all $k$ and $\gamma^*\in \Gamma^*$.
Another slightly more complicated choice is constructed with 
\be
 g_2: \R^d\ra  \R, \ 2\pi\Gamma^*-\mbox{anti-periodic}\ \  \mbox{i.e.} \ \ g_2(x+2\pi \gamma_i^*)=-g_2(x), \ \ \forall i=1,\dots,d,
\ee
and $\{\gamma_j^*\}_{j=1}^d$ is the basis of $\Gamma^*$. Then  $ f_2: \R^d\ra \R$ is defined as follows: for $j=1,\dots,d$, $e_j=\sum_{i=1}^d m_i(j)\gamma_i^*$, where $m_i(j)\in\Z$, and 
\be
f_2(x+2\pi e_j)=(-1)^{\sum_{i=1}^dm_i(j)}f_2(x), \ \ \forall x\in\R^d, \ \ \forall j=1,\dots,d.
\ee 
That is $f_2$ is $2\pi$-periodic or $2\pi$-anti-periodic in the direction $e_j$, depending on the components of the corresponding vector $\gamma_j^*$. Then, by construction, $b_2(k,p)=f_2(k)g_2(2p-k)$ satisfies our requirements and, moreover $b_2(k,p+2\pi \gamma_j^*/2)=-b_2(k,p)$.

Now, assume (\ref{point}) doesn't hold at some $p_0\in \T_\Gamma^d$. By a suitable choice of $g_1$ and $g_2$ as above,  we can construct a smooth $b(k,p)=b_1(k,p)+b_2(k,p)$ that is non-zero in a small neighborhood of $p_0$ only so that (\ref{difbar}) fails, which yields a contradiction.

Then one exploits the spectral decomposition (\ref{specdec}) and (\ref{conspecdec}) with $\Bra \delta_0\otimes\operatorname{Id}, R_0\Ket=1$ to deduce from the above that if $\|k\|$ is small enough
\bea
\lambda_1(k,p)&=&\lim_{n\ra\infty}\Big( \Bra  \delta_0\otimes\operatorname{Id}, \, {\widehat {M}(k,p)}^n
 R_0\Ket\Big)^{1/n}\nonumber\\
 &=&\lim_{n\ra\infty}\Big( \overline{\Bra  \delta_0\otimes\operatorname{Id}, \, {\widehat {M}(-k,p-k)}^n
 R_0\Ket}\Big)^{1/n}=\overline{\lambda_1(-k,p-k)}.
\eea
The result extends to complex $k$ by analyticity of $\lambda_1(\cdot,p)$ in $\cB_\kappa^d$.
\ep

\medskip

We now compute a second order expansion of $\lambda_1(k,p)=\tr (P_1(k,p)\widehat M(k,p)) $ around $k=0$, using the decomposition (\ref{decm})
\be
\widehat M(k,p)=\Sigma(k)S\widetilde Q,
\ee
where only the unitary map $\Sigma$ depends on $k$ (and $p$), as stressed in the notation.
We expand $\Sigma(k)$ as
\be
\Sigma(k)=\Sigma(0)+\Sigma_1(k)+\Sigma_2(k)+O_p(\|k\|^3),
\ee
where $\Sigma_j(k)$ is of order $j=1,2$ in $k$ and the remainder is $O_p(\|k\|^3)$, uniformly in $p\in \cT^d_\nu$. Explicitly,
\bea
&&(\Sigma_1(k)+\Sigma_2(k))\Psi(x,k,p;\zeta)=\\ \nonumber
&&\sum_{\tau,\tau'\in I^2_\pm} (ikr(\tau')-(kr(\tau'))^2/2) e^{ip(r(\tau)-r(\tau'))}P_\tau\Psi(x-r(\tau)+r(\tau'),k,p;\sigma_{-r(\tau')}\zeta)P_{\tau'}.
\eea
Then, in terms of the unperturbed reduced resolvent $S_p(z)$ defined for any $p\in \cT_\nu^d$ and $z$ in a neighborhood of 1, by
\be\label{defs1}
(\widehat M(0,p)-z)^{-1}=\frac{\Pi}{1-z}+S_p(z)
\ee
we have see \cite{k}, p. 79,
\bea
\lambda_1(k,p)&=&1+\tr (\Sigma_1(k)S\widetilde Q\Pi)\\ \nonumber
&+&\tr ( \Sigma_2(k)S\widetilde Q \Pi-\Sigma_1(k)S\widetilde QS_p(1)\Sigma_1(k)S\widetilde Q\Pi)+O_p(\|k\|^3).
\eea
Explicit computations making use of $S\widetilde Q\widehat \Psi_0=\widehat\Psi_0$, $ S\widetilde QS_p(1)=\Sigma(0)^{-1}(\I-\Pi + S_p(1))$ and 
\be
(\Sigma(0)^{-1}\Phi)(x,\eta)=\sum_{(\tau,\tau')\in I^2_\pm}e^{-ip(r(\tau)-r(\tau'))}P_\tau
\Phi((x+r(\tau)-r(\tau'), \sigma_{r(\tau')})P_{\tau'}
\ee
yield
\begin{lem}\label{secord}
For all $p\in \cT^d_\nu$ and $k\in \cB(0,\kappa)$, there exists a symmetric matrix $\D(p)\in M_{d}(\C)$  such that
\bea\label{difmat}
&&\lambda_1(k,p)=1+\frac{i}{2d}\sum_{\tau\in I_\pm}kr(\tau)+O_p(\|k\|^3)\nonumber \\
&&\hspace{.5cm}+\frac{1}{2d}\left(\sum_{\tau\in I_\pm}\frac{(kr(\tau))^2}{2}+\sum_{\tau,\tau'\in I_\pm}(kr(\tau))(kr(\tau'))\left\{\Bra\delta_0\otimes P_{\tau'} | (S_p(1)\delta_0\otimes P_\tau)\Ket_{l^2}-\frac{1}{2d}\right\}\right)\nonumber \\
 &&\hspace{1.5cm}\equiv1+\frac{i}{2d}\sum_{\tau\in I_\pm}kr(\tau)-\frac 12 \bra k | \D(p) k\ket+ O_p(\|k\|^3).
\eea
The map $p \mapsto \D(p)$ is analytic on $\cT_\nu^d$; when $p\in \T_\Gamma^d$, $\D(p)\in M_{2d}(\R)$ is non-negative and $\D(p)_{i,j}=\frac{\partial^2}{\partial k_i \partial k_j}\lambda(0,p)$, $i,j \in \{1,2,\dots, d\}$. Moreover,  $O_p(\|k\|^3)$ is uniform in $p\in \cT_\nu^d$.
\end{lem}
\proof Existence and analyticity in $p$ of $\D(p)$ follow from analyticity of $\lambda_1$ in $y$ and analyticity of $S_p(1)$ in $p$, see (\ref{defs1}). Since $\D(p)_{i,j}=\frac{\partial^2}{\partial k_i \partial k_j}\lambda(0,p)$,  the matrix is symmetric.
For $p\in\T_\Gamma^d$, Lemma \ref{rvl} implies that $\lambda(0,p)$ is real valued, hence the matrix elements $\D(p)$ for $p\in\T_\Gamma^d$ are real as well. Finally, (\ref{bddspec}) implies that $\bra k | \D(p) k\ket\geq 0$ for all $k\in\T^d$. 
\ep

\medskip

As a consequence of the spectral analysis above, it follows exactly as in \cite{J1}, that 
\begin{prop}\label{tecprop}
Under assumption {\bf S}, uniformly in $p\in\cT^d_\nu$, in $k$ in compact sets of $\C^d$ and in $t$ in compact sets of $\R_+^*$,
\bea
&&\lim_{n\ra\infty}\widehat M(k/{n},p)^{[tn]}=e^{i{t} y \overline{r}}\Pi,\\
&&\lim_{n\ra\infty}\widehat M(k/\sqrt{n},p)^{[tn]}e^{-i[tn]\overline{r}y/\sqrt{n}}=e^{-\frac{t}{2}\bra y | \D(p) y\ket}\Pi.
\eea
\end{prop}

\section{Diffusion Properties}\label{diff}
These technical results lead to the main results of this section which are the existence of a diffusion matrix  and central limit type behaviors in the diffusive scaling, as in \cite{J1}, with the same proofs, that we don't need to repeat.

\medskip

Let  $\cN(0,\Sigma)$ denote the centered normal law in $\R^d$ with positive definite covariance matrix $\Sigma$ and let us write  $X^{\omega}\simeq \cN(0,\Sigma)$ a random vector $X^\omega\in\R^d$ with distribution $\cN(0,\Sigma)$. The superscript $\omega$ can be thought of as a vector in $\R^d$ such that for any Borel set $A\subset {\mathbb R}^d$
\be
\P(X^\omega\in A)=\frac{1}{(2\pi)^{d/2}\sqrt{\det (\Sigma)}}\int_{A}  e^{-\frac{1}{2}\bra \omega|\Sigma^{-1} \omega\ket} d\omega.
\ee
The corresponding characteristic function is $\Phi^{\cN}(y)=\E(e^{iyX^{\omega}})=e^{-\frac12\bra y | \Sigma y\ket}$.

\medskip 

The first result concerning the asymptotics of the random variable $X_n$ reads as follows for an initial density matrix of the form  $\rho_0=|\ffi_0\ket\bra\ffi_0|\otimes |0\ket\bra 0|$ :

\begin{thm} \label{cf} 
Under Assumption  {\bf C} and {\bf S}, uniformly in $y$ in compact sets of $\C^d$ and in $t$ in compact sets of $\R_+^*$,
\bea
&&\lim_{n\ra\infty}\Phi^{\ffi_0}_{[tn]}(y/ n)=e^{i{t} y  \overline{r}}\\
&&\lim_{n\ra\infty}e^{-i[tn]\frac{\overline{r}y}{\sqrt{n}}}\Phi^{\ffi_0}_{[tn]}(y/\sqrt n)=\int_{\T_\Gamma^d} e^{-\frac{t}{2}\bra y | \D(p) y\ket}\, d{\tilde p},
\eea
where the right hand side admits an analytic continuation in $(t,y)\in \C\times \C^d$.
In particular, for any $(i,j)\in \{1, 2, \dots, d\}^2$, 
\bea
&& \lim_{n\ra\infty}\frac{\bra X_i\ket_{\psi_0}(n)}{n}=\overline{r}_i \\
&&\lim_{n\ra\infty}\frac{\bra (X-n\overline{r})_i (X-n\overline{r})_j \ket_{\psi_0}(n)}{n}=\int_{\T_\Gamma^d} \D_{i\,j}(p)\, d{\tilde p}.
\eea
If $\D(p)=\D>0$ is independent of $p\in \T_\Gamma^d$, then, for any initial vector $\Psi_0=\ffi_0\otimes |0\ket$,  we have as $n\ra\infty$, with convergence in law,
 \bea
 \frac{X_n-n\overline r}{\sqrt n}&{\cD \atop \longrightarrow }&\ X^\omega\simeq \cN(0,\D).
 \eea

\end{thm} 
\begin{rem} 
We will call {\em diffusion matrices} both $\D(p)$ and 
$
\D=\int_{\T_\Gamma^d} \D(p)\, d{\tilde p}.
$
\end{rem}
\begin{rem}
We prove below that a central limit theorem for $X_n$ may hold in cases where $\D$ depends on $p$, see Theorem \ref{bryc}.
\end{rem}

For initial conditions corresponding to a density matrix $\rho_0$, we have
\begin{cor}\label{cdiffeq}
Under Assumptions {\bf C }, {\bf S} and {\bf R} for the observable $X^2$, we have for any $t\geq0$, $y\in\C^d$,
\bea
\lim_{n\ra\infty}\Phi^{\rho_0}_{[tn]}(y/n)&=&e^{i{t} y  \overline{r}}, \\
\lim_{n\ra\infty} e^{-i[tn]\frac{\overline{r}y}{\sqrt{n}}}\Phi^{\rho_0}_{[tn]}(y/\sqrt n)&=&
\int_{\T_\Gamma^d} e^{-\frac{t}{2}\bra y | \D(p) y\ket}\Bra\Psi_1|\Pi \widehat \rho_0 (\cdot, 0, p, \cdot)\Ket_{L^2(B_\Gamma\times \T^d\times\Omega;M_{2d}(\C))}d{\tilde p}
\nonumber\\
&=&
\int_{\T_\Gamma^d} e^{-\frac{t}{2}\bra y | \D(p) y\ket}\, 
{\em \tr} (\widehat \rho_0)(0, 0, p)
d{\tilde p}
\eea
where, see (\ref{ideta}),
\be
\widehat \rho_0(0,0,p)= \sum_{{\xi\in\Z^d}\atop{\zeta\in \Gamma}}e^{-ip\zeta}
\rho_0(\xi-\zeta, \xi).
\ee
Also, for any $(i,j)\in \{1, 2, \dots, d\}^2$, 
\bea
&&\lim_{n\ra\infty}\frac{\bra X_i\ket_{\rho_0}(n)}{n}=\overline{r}_i,\\
&&\lim_{n\ra\infty}\frac{\bra (X-n\overline{r})_i(X-n\overline{r})_j\ket_{\rho_0}(n)}{n}=
\int_{\T_\Gamma^d}  \D_{i\,j}(p)\, 
{\em \tr} (\widehat \rho_0)(0, 0, p)
d{\tilde p}.
\eea
\end{cor}
From Corollary \ref{cdiffeq}, and Theorem \ref{cf}, we gather that the characteristic function of the centered variable $X_n-n\overline r$ in the diffusive scaling $T=nt$, $Y=y/\sqrt{n}$, where $n\ra\infty$, converges to 
\be
\int_{\T_\Gamma^d} 
\cF\left(\frac{e^{-\frac{1}{2t}\bra \cdot | \D^{-1}(p) \cdot \ket} }{(t 2\pi)^{d/2}\sqrt{\det \D(p)}}\right)(y)
\, { \tr} (\widehat \rho_0)(0, 0, p) d{\tilde p},
\ee
where the function under the Fourier transform symbol $\cF$ is a solution to the diffusion equation
\be\label{diffusion}
\frac{\partial \ffi}{\partial t}=\frac12\sum_{i,j=1}^d\D_{ij}(p)\frac{\partial^2\ffi}{\partial_{x_i}\partial_{x_j}}.
\ee
As explained in \cite{ks}, \cite{hks}, it follows that the position space density $w_k([nt])\delta(\sqrt n x-k)$ converges in the sense of distributions to a superposition of solutions to the diffusion equations (\ref{diffusion}) as $n\ra\infty$.

\section{Moderate Deviations}\label{moder}

It is shown in \cite{J1} that the spectral properties of the matrix $\widehat M(k,p)$ proven in Section \ref{sp} allow us to obtain further results on the behavior with $n$ of the distribution of the random variable $X_n$ defined by (\ref{wW}) with localized initial condition $\rho_0=|\ffi_0\ket\bra\ffi_0|\otimes |0\ket\bra 0|$, corresponding to the vector $R_0\in l^2(B_\Gamma\times\Omega; M_{2d}(\C))$, see (\ref{rz}). This section is devoted to establishing some moderate deviations results on the centered random variable $X_n-n\overline r$. Again, since all proofs are identical to those given in \cite{J1}, we merely state the results.

\medskip

Moderate deviations results depend on asymptotic behaviors in different regimes of the logarithmic generating function of $X_n-n\overline r$ defined for $y\in\R^d$ by
\be
\Lambda_n(y)=\ln(\E_{w(n)}(e^{y(X_n-n\overline r)}))\in (-\infty, \infty].
\ee
This function $\Lambda_n$ is convex and $\Lambda_n(0)=0$.

Let $\{a_n\}_{n\in \N}$ be a positive valued sequence such that 
\be\label{an}
\lim_{n\ra\infty}a_n=\infty, \ \mbox{ and } \ \ \lim_{n\ra\infty}a_n/n=0.
\ee Define $Y_n=(X_n-n\overline r)/\sqrt{n a_n}$ and, for any $y\in \R^d$, let $\tilde \Lambda_n(y)=\ln(\E_{w(n)}(e^{yY_n}))$ be the logarithmic generating function of $Y_n$. 

\begin{prop} \label{51} Assume {\bf C} and {\bf S} and further suppose $\D(p)>0$ for  all $p\in \T^d$. Let $y\in \R^d\setminus \{0\}$ and assume the real analytic map  $\T^d\ni p\mapsto \bra y | \D(p)y\ket\in\R_*^+$ is either constant or admits a finite set $\{p_j(y)\}_{j=1, \cdots, J}$ of non-degenerate maximum points in $\T^d$.
Then, for any $y\in\R^d$, 
\bea
\lim_{n\ra\infty}\frac{1}{a_n}\tilde \Lambda_n(a_n y)=\frac12 \bra y | \D(p_1(y))y\ket
\eea
which is a smooth convex function of $y$.
\end{prop}

Let us introduce a few more definitions and notations. A {\em rate function} $I$ is a lower semicontinuous map from $\R^d$ to $[0,\infty]$ s.t. for all $\alpha\geq 0$, the level sets $\{x\ | \ I(x)\leq \alpha\}$ are closed. When the level sets are compact, the rate function $I$ is called {\em good}. For any set $\Gamma\subset \R^d$, $\Gamma^0$ denotes the interior of $\Gamma$, while $\overline{\Gamma}$ denotes its closure.

As a direct consequence of G\"artner-Ellis Theorem, see \cite{dz} Section 2.3, we get
\begin{thm}\label{md} Define $\Lambda^*(x)=\sup_{y\in \R^d}\left(\bra y|x\ket -\frac12\bra y | \D(p_1(y))y\ket\right)$, for all $x\in \R^d$. Then, $\Lambda^*$ is a good rate function and, for any positive valued sequence $\{a_n\}_{n\in \N}$ satisfying (\ref{an}) and  all Borel set $\Gamma\subset \R^d$
\bea
-\inf_{x\in \Gamma^0}\Lambda^*(x)&\leq &\liminf_{n\ra\infty}\frac{1}{a_n}\ln (\P((X_n-n\overline r)\in \sqrt{na_n}\, \Gamma))\nonumber\\
&\leq&\limsup \frac{1}{a_n}\ln (\P((X_n-n\overline r)\in \sqrt{na_n}\, \Gamma))\leq -\inf_{x\in \overline{\Gamma}}\Lambda^*(x).
\eea
\end{thm}
\begin{rem} As a particular case, when $\D(p)=\D>0$ is constant, we get
\be
\Lambda^*(x)=\frac12\bra x| \D^{-1}x\ket.
\ee
\end{rem}
\begin{rem}
Specializing the sequence $\{a_n\}_{n\in \N}$ to a power law, {\it i.e.} taking $a_n=n^\alpha$, we can express the content of Theorem \ref{md} in an informal way as follows. For $0<\alpha<1$, 
\be
\P((X_n-n\overline r)\in n^{(\alpha +1)/2}\, \Gamma)\simeq e^{-n^\alpha \inf_{x\in {\Gamma}}\Lambda^*(x)}.
\ee
For $\alpha$ close to zero, we get results compatible with the central limit theorem and for $\alpha$ close to one, we get results compatible with those obtained from a large deviation principle.
\end{rem}


\section{Large Deviations}\label{LargeDev}

In this section, we push further  the analysis of the large $n$ behavior distribution of the random variable $X_n$ (defined by (\ref{wW}) with localized initial condition $\rho_0=|\ffi_0\ket\bra\ffi_0|\otimes |0\ket\bra 0|$) by proving large deviations estimates  and a central limit theorem under stronger assumptions on the spectral properties of the matrix $\widehat M(k,p) $.
\medskip

We change scales and define for $n\in\N^*$ and $y\in\R^d$ a rescaled random variable and the corresponding convex logarithmic generating  function
\be\label{yl}
Y_n=\frac{X_n-n\bar r}{n}  \ \ \mbox{and} \ \ \tilde\Lambda_n(y)=\ln \E_{w(n)}(e^{yY_n})\in (-\infty,\infty].
\ee
Because of the new scale $n$, the existence of $\lim_{n\ra\infty }\frac{\tilde \Lambda(ny)}{n}=\lim_{n\ra\infty } \frac{\ln \E_{w(n)}(e^{yX_n})}{n}-y\bar r$ is not granted for all $y$, by contrast with the previous section. However, because $\|Y_n\|\leq c_0$, for some $c_0<\infty$, we have for any $y\in\C^d$, and {\em a fortiori} for any $y\in\R^d$,
\be\label{limsup}
\frac{ | \tilde \Lambda(ny) |}{n}\leq \|y\|c_0.
\ee 
Moreover, as the next Proposition states, the limit exists for $\|y\|$ small enough, under more global, yet reasonable, hypotheses:
\begin{prop} \label{lapl} Let $y\in \R^d\cap B(0,\kappa)$ be fixed and assume the function $\T_\Gamma^d\ni p\mapsto|\lambda_1(-iy,p)|$ is either constant or admits a finite set of non-degenerate global maxima  $\{p_j(y)\}_{j=1,\dots, N}$ in $\T_\Gamma^d$. Further assume $\nabla_{p}\lambda_1(-iy,p_j(y))=0$, for all $j=1,\dots, N$.  Then, for $\kappa>0$ small enough,
\be\label{rhs}
\lim_{n\ra\infty }\frac{\tilde \Lambda(ny)}{n}=-y\bar r+\ln(\lambda_1(-iy,p_1(y)))
\ee
\end{prop}
is a smooth real valued convex function of $y\in B(0,\kappa)\cap\R^d.$
\begin{rems}
i) In case $\lambda_1(-iy,p)\equiv \lambda_1(-iy,0)$ is independent of $p\in\T^d$, the right hand side of (\ref{rhs}) equals $-y\bar r+\ln(\lambda_1(-iy,0))$.\\
ii) The assumption $\nabla_{p}\lambda_1(-iy,p_j(y))=0$ may be too strong to deal with certain cases. However, if it does not hold, in which case $\nabla_{p}\lambda_1(-iy,p_j(y))\in i\R^d$, the asymptotics of the integral that yields $\tilde \Lambda(ny)/n$ is out of reach of a steepest descent method without further information on the behavior of $\lambda_1(-iy,p)$ for $p$ away from $\T_\Gamma^p$.
\end{rems}
\medskip

The proof is a straightforward alteration of that of Proposition \ref{51}, based on Laplace's method to evaluate the asymptotics of the integral
\bea\label{64}
 \E_{w(n)}(e^{nyY_n})&=&e^{-ny\bar r}\int_{\T_\Gamma^d} \big\bra\Psi_1| \widehat M^n(-iy,p)R_0\big\ket_{l^2(B_\Gamma\times\Omega, M_{2d}(\C))} d\tilde p\\ \nonumber
&=& e^{-ny\bar r}\int_{\T_\Gamma^d} e^{n\ln(\lambda_1(-iy,p))}\big\bra\Psi_1|  P(-iy,p)R_0\big\ket_{l^2(B_\Gamma\times\Omega, M_{2d}(\C))} d\tilde p+O_p(e^{-n\gamma }), 
\eea
where $ \gamma >0$ and the prefactor is non-zero, due to the smallness of $\|y\|$. For completeness, we briefly recall the argument in case there is only one maximum at $p_1\in \T_\Gamma^d$. Dropping the variable $y$ in the notation and writing $\ln(\lambda_1(p))=a(p)+ib(p)$, $\cP(p)=\big\bra\Psi_1|  P(-iy,p)R_0\big\ket_{l^2(B_\Gamma\times\Omega, M_{2d}(\C))} $ we have in a neighborhood of $p_1\in\T_\Gamma^d$ determined by $\nabla a(p_1)=0$ and  $D^2a(p_1)<0$
\bea\label{asym}
&&e^{n\ln(\lambda_1(p))}\cP(p)=e^{n \ln(\lambda_1(p_1))}\cP(p_1)\times\\ \nonumber
&&\hspace{.9cm}\times e^{in\bra \nabla b(p_1)(p-p_1)\ket}e^{n \bra (p-p_1)| (D^2a(p_1)+iD^2b(p_1))(p-p_1)\ket/2 }e^{nO(\|p-p_1\|^3)}(1+O(\|p-p_1\|)).
\eea
Making use of $D^2a(p_1)<0$, we can restrict the integration range in (\ref{64}) to $B(p_1,\mu(n))\subset \R^d$, with $1/\sqrt{n}\ll \mu(n)\ll 1/n^{1/3}$, at the cost of an error of order $e^{-n\mu(n)^2c}$, for some $c>0$, so that we are lead to
\be\label{restint}
\int_{B(0,\mu(n))}e^{in\bra\nabla b(p_1)|p\ket}e^{n \bra p| (D^2a(p_1)+iD^2b(p_1)) p\ket/2 }dp\, (1+O(n\mu(n)^3)+O(\mu(n))).
\ee
When $\nabla b(p_1)\neq 0$, the analysis of the large $n$ behavior of (\ref{64}) and (\ref{restint}) requires global informations about the analytic properties of $\lambda_1$ for $p$ far from the real set $\T_\Gamma^d$, hence we require $\nabla b(p_1)=0$.
Since $\lambda_1=1+O(\|y\|)\neq 0$, we have
\bea
\nabla a(p_1)&=&0 \ \ \Leftrightarrow\ \ \Re \lambda_1(p_1)\nabla \Re \lambda_1(p_1)+\Im \lambda_1(p_1)\nabla \Im \lambda_1(p_1)=0\\
\nabla b(p_1)&=&\nabla \arg(\lambda_1(p))=\frac{\nabla \Im \lambda_1(p_1)}{\Re \lambda_1(p_1)},
\eea
so that the hypothesis $\nabla b(p_1)\in\R^d$ implies $\nabla \lambda_1(p_1)=0$.
Now, at the cost of another error of order  $e^{-n\mu(n)^2c}$, (\ref{restint}) equals
\be
\int_{\R^d}e^{n \bra p| (D^2a(p_1)+iD^2b(p_1)) p\ket/2 }dp\, (1+O(n\mu(n)^3)+O(\mu(n))) + O(e^{-n\mu(n)^2c}),
\ee
where a Gaussian integral yields
\be
\int_{\R^d}e^{n \bra p| (D^2a(p_1)+iD^2b(p_1)) p\ket/2 }dp=\frac{G}{n^{d/2}}, \ \ \mbox{where} \ \ G=\frac{(2\pi)^{d/2}}{\sqrt{\det(-D^2a(p_1)-iD^2b(p_1))}}.
\ee
Altogether, we get
\bea
\frac{\ln( \E_{w(n)}(e^{nyY_n}))}{n}&=&\ln(\lambda_1(-iy,p))\\ \nonumber
&+&\frac{1}{n}\ln\Big(\frac{G\cP(p_1)}{n^{d/2}}(1+O(n\mu(n)^3)+O(\mu(n))+O(e^{-n\mu(n)^2c})\Big),
\eea
which yields the result in the limit $n\ra\infty$. \ep

\medskip

We set for all $y\in\R^d$
\be\label{ll}
\overline\Lambda(y)=\limsup_{n\ra\infty}\frac{\tilde \Lambda(ny)}{n}\in (-\infty,\infty),
\ee
which is convex, finite everywhere and bounded by $c_0\|y\|$. Moreover, for $\|y\|<\kappa$, $\overline\Lambda(y)$ equals the right hand side of (\ref{rhs}) and is thus smooth, and $\overline\Lambda(0)=\Lambda(0)=0$.
Let us consider the Legendre transform of $\overline \Lambda$ 
\be\label{lsl}
\overline\Lambda^*(x)=\sup_{y\in \R^d}\left(\bra y|x\ket -\overline \Lambda(y)\right)\geq 0, \  \mbox{for all }\ \ x\in \R^d.
\ee
We are now in a position to state our large deviations results via G\"artner-Ellis Theorem.

\begin{thm} Assume the hypothese of Proposition \ref{lapl}
Let $\overline\Lambda$ and $\overline\Lambda^*$ be defined by (\ref{ll}) and (\ref{lsl}). Further assume $\overline\Lambda$ is strictly convex in  neighborhood of the origin. Then $\overline\Lambda^*$ is a good rate function and there exists $\eta>0$ such that for any $\Gamma\in\R^d$
\bea\label{ubge}
&&\limsup \frac{1}{n}\ln (\P((X_n-n\overline r)\in n\, \Gamma))\leq -\inf_{x\in \overline{\Gamma}}\overline\Lambda^*(x)\\ \label{lbge}
&&\liminf_{n\ra\infty}\frac{1}{n}\ln (\P((X_n-n\overline r)\in n\, \Gamma))\geq -\inf_{x\in \Gamma^0\cap B(0,\eta)}\overline\Lambda^*(x).
\eea
\end{thm}
\proof
Exercise 2.3.25, p. 54 in \cite{dz},  shows that since $\overline\Lambda$ is finite on $\R^d$ then $\overline\Lambda^*$ is a good rate function and that (\ref{ubge}) holds. 

To show that (\ref{lbge}) holds, we invoke Baldi's Theorem, Thm 4.5.20 in \cite{dz}. First, Exercise 4.1.10 of \cite{dz}, point c), shows that the law of $Y_n$ is exponentially tight, as a consequence of $\overline\Lambda^*$ being a good rate function and (\ref{ubge}) holding true. Then, by Exercise 2.3.25 still,   if $x=\nabla \overline\Lambda(y)=\nabla \Lambda(y)$ for some $y\in B(0,\kappa)$, then $x\in {\cal F}$, where ${\cal F}$  is the set of {\it exposed point} for $\overline\Lambda^*$ with {\it exposing hyperplane} $y$. Let us recall that this means that for all $z\neq x$, $yx-\overline\Lambda^*(x)>yz-\overline\Lambda^*(z)$. Now, since $\overline\Lambda$ is strictly convex at the origin, its Hessian at zero is positive definite and $\nabla \overline\Lambda(0)=0$. It thus follows from the implicit function theorem that for some $\eta>0$, the map $y\ra \nabla \overline\Lambda(y)$ is a bijection with range $B(0,\eta)$. Hence $B(0,\eta)$ is included in the set of exposed points for $\overline\Lambda^*$.  Also, the corresponding set of exposing hyperplanes belongs to $B(0,\kappa)$, where $\Lambda$ coincides with $\overline\Lambda$, which is finite everywhere. Hence, all hypotheses of Baldi's Theorem are met, so that  (\ref{lbge}) holds.
\ep
\\

Another direct consequence of Proposition \ref{lapl} together with (\ref{limsup}) is a central limit theorem for $X_n$, as proven by Bryc, \cite{b}. A vector valued version of Bryc's Theorem suited for our purpose can be found in \cite{jopp}.
\begin{thm}\label{bryc}
Under the assumptions of Proposition \ref{lapl}, we have, with convergence in law,
\be
\frac{X_n-n\bar r}{n^{1/2}}\ {\longrightarrow}\ {\cal N}(0,\D)
\ee
where $\D_{i,j}=\frac{\partial^2}{\partial_{y_i}\partial_{y_j}}\Lambda(0)\geq 0$.
\end{thm}
\begin{rem} 
The results of this section carry over to the cases considered in \cite{J1}, see also Section \ref{uncorr}.
\end{rem}


\bigskip

\section{Example}\label{11}

Let us consider here a fairly general situation in which the spectral hypotheses we need can be checked explicitly.

\medskip

We work in $\Z^d$ and consider a model characterized by a representation of $\Z^d$, $x\mapsto \sigma_x$ of measure preserving maps, a jump function $r:I_\pm\ra\Z^d$ such that
\be\label{ajf}
r(\tau)-r(\tau')\in \Gamma, \ \ \forall \tau,\tau'\in I_\pm,
\ee
a kernel $\P$ with identical lines 
\be\label{akp}
\P(\eta,\zeta)=\P(\zeta), \ \ \forall \eta\in \Omega,
\ee
and a set of unitary matrices $\{\eta\}_{\eta\in \Omega}$ with trivial commutant
\be\label{aum}
\{\eta\}_{\eta\in \Omega}'=\{c\un, c\in\C\}.
\ee
This implies that the corresponding stationary distribution is $p(\zeta)=\P(\zeta)$.
We address the simplicity of the eigenvalue 1 of ${\widehat M}(0,p)|_{\cal I}$, see Remark \ref{mresi}.

\begin{prop} Under assumptions (\ref{ajf}), (\ref{akp}) and (\ref{aum}), ${\widehat M}(k,p)|_{\cal I}$ is independent of $p$ and ${\widehat M}(0,p)|_{\cal I}$ admits 1 as a simple eigenvalue.
\end{prop}
\proof
The simplicity of the eigenvalue 1 of ${\widehat M}(0,p)|_{\cal I}$ is equivalent to the simplicity of the eigenvalue 1 of ${\widehat M}(0,p)^*|_{\cal I}$. 

We first observe that ${\widehat M}(k,p)^*$ leaves the subspace 
\be
{\cal J}\equiv \mbox{span}\{\delta_0\otimes A\, |\, A:\Omega\ra M_{2d}(\C) \ \mbox{is constant}\}
\ee
invariant:
\bea
&&(\widehat M(k,p)^*\delta_0\otimes A)(x,\eta)
\\ \nonumber
&&=\sum_{\tau,\tau'\in I_\pm\atop{\zeta\in\Omega}} p(\zeta)e^{-ikr(\tau')} e^{-ip(r(\tau)-r(\tau'))}(\sigma_{x}\zeta)^*  \delta_0\left(x+r(\tau)-r(\tau')\right)P_\tau A P_{\tau'} \zeta\\ \nonumber
&&=e^{ipx}\delta_0\left(x\right)\sum_{\tau,\tau'\in I_\pm\atop{\zeta\in\Omega}}  p(\zeta)
(\sigma_{x}\zeta)^* P_\tau A P_{\tau'}e^{-ikr(\tau')} \zeta
\\ \nonumber
&&=\delta_0(x)\sum_{\zeta\in\Omega} p(\zeta)\zeta^* AU(k)\zeta,
\eea
where $U(k)= \sum_{\tau'\in I_\pm}P_{\tau'}e^{-ikr(\tau')} $. Hence we  have $
{\cal I}\subset \cal J$
and ${\widehat M}(k,p)^*|_{\cal J}$ is independent of $p\in \T_\Gamma^d$. Thus we can consider ${\widehat M}(0,p)^*|_{\cal J}$. 

Note that $U(0)=\un$ and that ${\widehat M}(0,p)^*|_{\cal J}\delta_0\otimes A=\delta_0\otimes A$ is equivalent to $\cM(A)=A$ where
\be
\cM(A):= \sum_{\zeta\in\Omega} p(\zeta)\zeta^* A \zeta, \ \ \ \forall A\in M_{2d}(\C).
\ee
With the scalar product $\bra A|B\ket=\tr (A^*B)$ on $M_{2d}(\C)$ we have
\bea
\|\cM(A)\|^2=\sum_{(\zeta, \eta)\in\Omega^2} p(\zeta)p(\eta)\bra \eta^*A\eta|\zeta^* A \zeta\ket,
\eea
where $|\bra \eta^*A\eta|\zeta^* A \zeta\ket |\leq \|A\|^2$, with equality if and only if 
$ \eta^*A\eta=e^{i\theta_{\eta \zeta}}\zeta^* A \zeta$, for some $\theta_{\eta \zeta}\in \R$. Hence $\|\cM(A)\|=\|A\|$ if and only if $ \eta^*A\eta=\zeta^* A \zeta$, for all $\eta, \zeta$. Thus, any invariant matrix under $\cM$ satisfies 
\be
\cM(A)= \eta^*A\eta=A, \ \ \forall \eta\in \Omega.
\ee
Since the commutant of $\{\eta\}_{\eta\in\Omega}$ is assumed to be reduced to $c\un$, $c\in\C$, we get the result. \ep

\section{Examples of diffusive  random dynamics}\label{four}

In this section we consider a specific example of measure $d\mu$ on $U(2d)$, the set of coin matrices, for which we can prove convergence results on the random quantum dynamical system associated with (\ref{racm}) for large times. 
This example is a generalization of the example  considered in \cite{J1} for site-independent coin matrices. While the following results hold for vector and  density matrix initial conditions, we only consider here the vector case, for shortness.

\subsection{Permutation matrices}

We start by recalling a few deterministic facts.
Let ${\mathfrak S}_{2d}$ be the set of permutations of the $2d$ elements of $I_\pm=\{\pm 1, \pm 2, \dots, \pm d\}$. For $\pi\in {\mathfrak S}_{2d}$,  define 
\be\label{defmatper}
C(\pi)=\sum_{\tau\in I_\pm}|\pi(\tau)\ket\bra \tau | \in U(2d) \ \ \mbox{so that }\ C_{\sigma \tau}(\pi)=\delta_{\sigma, \pi(\tau)},
\ee
and
$C(\pi)$ is a permutation matrix associated with $\pi$. Note the elementary properties: 
for any $\pi, \sigma \in {\mathfrak S}_{2d}$, 
\bea
C(\I)=\I, \ \ C^*(\pi)=C^T(\pi)=C(\pi^{-1}), \ \ C(\pi)C(\sigma)=C(\pi\sigma).
\eea

\medskip

The matrices $C(\pi)$ allow for explicit computations of the relevant quantities introduced in Section \ref{dets}. Given a sequence $\{C_j=C(\pi_j)\}_{j=1,\dots, n}$ of such matrices, a direct computation shows that with the definition $\tau_j=\pi_j (\tau_{j-1})$, $J^0_k(n)$ takes the form
\bea
J^0_{k}(n)&=&\sum_{\tau_1\in I_\pm \ s.t. \atop  \sum_{j=1}^nr(\tau_j)=k}
|\tau_n\ket\bra \pi_1^{-1}(\tau_1)|,
\eea
and $J^0_k(n)=0$, if $\sum_{j=1}^nr(\tau_j)\neq k$. 

Consequently, the non-zero probabilities $W_k(n)$ on $\Z^d$ read for any normalized internal state vector $\ffi_0$.
\bea
W_k^{\ffi_0}(n)&=&\|J^0_k(n)\ffi_0\|^2= \sum_{\tau_1\in I_\pm \ s.t. \atop  \sum_{j=1}^nr(\tau_j)=k}|\bra \pi_1^{-1}(\tau_1)|\ffi_0\ket|^2.
\eea
Moreover, with $\tau_1=\pi_1(\tau_0)$ we get
\be
\ffi_0=\sum_{\tau_0 \in I_\pm}a_{\tau_0} |\tau_0\ket \ \ \Rightarrow \ \ |\bra \pi_1^{-1}(\tau_1)|\ffi_0\ket|^2=\sum_{\tau_0\in\I_\pm}|a_{\tau_0}|^2\delta_{\tau_1, \pi_1(\tau_0)}.
\ee
Hence $W_k^{\ffi_0}(n)= \sum_{\tau_0\in I_\pm }
|a_{\tau_0}|^2\delta_{\sum_{j=1}^nr(\tau_j), k}$ \ so that for $F=\I\otimes f$ and $\psi_0=\ffi_0\otimes |0\ket\bra 0|$
\be
\bra F\ket_{\psi_0}(n)= \sum_{k\in \Z^d}W_k^{\ffi_0}(n)f(k)= \sum_{\tau_0\in I_\pm }|a_{\tau_0}|^2 f(\sum_{j=1}^nr(\tau_j)).
\ee
\begin{rems}
In other words, given a set of $n$ permutations, there is no more quantum randomness in the variable $X_n$, except in the initial state. \\
If one generalizes the set of matrices by adding phases to the matrix elements of the permutation matrices, it does not change the probability distribution $\{W_k^{\ffi_0}(n)\}_{k\in\Z^d}$, see \cite{J1}.
\end{rems}
Therefore the characteristic functions take the form
\begin{cor} With $\tau_j=(\pi_j \pi_{j-1}\cdots \pi_1)(\tau_0)$, for $j=1,\dots, n$,
\bea\label{carran}
\Phi^{\ffi_0}_n(y)&=&\sum_{\tau_0\in I_\pm }e^{iy \sum_{j=1}^nr(\tau_j)}|a_{\tau_0}|^2
\eea
\end{cor}
The dynamical information is contained in the sum $S_n=\sum_{j=1}^nr(\tau_j)$ which appears in the phase. The next section is devoted to its study, in the random version of this model where the coin matrices are random variables with values in   $\{C(\pi), \, \pi\in {\mathfrak S}_{2d}\}$ distributed according to (\ref{racm})

\subsection{Random Setup}

\medskip

We consider that the permutation matrices are given by the process  defined by  (\ref{racm}) and we identify $C(\pi)$ and $\pi$:

\medskip 
\noindent {\bf Assumption  $\widetilde{\bf M}$:}\\
Let $\{\omega(n)\}_{n\in\N}$ be a finite state space Markov chain on $\Omega\subset {\mathfrak S}_{2d}$ with transition matrix ${\mathbb P}$ and stationary  initial distribution $p$ and a representation $\sigma$ of $\Z^d$ of the form $x\mapsto \sigma_x$ where for each $x\in \Z^d$, $\sigma_x:\Omega\ra\Omega$ with $\Omega \subset {\mathfrak S}_{2d}$, is a measure preserving bijection. We set $C_n^\omega(x)=\sigma_x(\omega(n))$ with $C_n^\omega(0)=\omega(n)$. 
 \medskip

We have that for every $x\in\Z^d$, the  set of random matrices/permutations $\{\pi_n^\omega(x))\}_{n\in\N}=\{\sigma_x(\omega(n))\}_{n\in\N}$, with $\omega(n)\in\Omega\subset { \mathfrak S}_{2d}$, the Markov chain.

\medskip

Given a set of random permutation matrices as above, we start at time zero on site $0\in\Z^d$, with initial vector $|\tau_0\ket\otimes |0\ket$.  The dynamics induced by  the permutation matrices sends this state at time $n\geq 1$ to the state $|\sum_{s=1}^nr(\tau_s)\ket\otimes |\tau_n\ket$, 
where $\tau_j=\sigma_{\sum_{s=1}^{j-1}r(\tau_s)}(\omega(j))\tau_{j-1}.$

Hence, in view of (\ref{carran}),  we introduce the random variables $S_n(\overline \omega)=\sum_{j=1}^nr(\tau_j(\overline\omega))\in \Z^d$ and  $r(\tau_j(\overline\omega))$ where
$\tau_j(\overline \omega)$ is defined for $j=1,\dots,n$ by
\be\label{process}
\tau_1(\overline \omega)=\sigma_{0}(\omega(1))\tau_{0},\ \ \ \tau_j(\overline \omega)=\sigma_{\sum_{s=1}^{j-1}r(\tau_{s}(\overline \omega))}(\omega(j))\tau_{j-1}(\overline \omega),
\ee
for a given $\tau_0$. Note that $ \tau_j(\overline \omega)= \tau_j((\omega(j), \omega(j-1), \dots , \omega(1))$. They have the following properties
\begin{lem}\label{nonstat} Let $\ffi_0=\sum_{\tau_0}a_{\tau_0}|\tau_0\ket$ be the initial vector, and assume {$\widetilde {\bf M}$} holds true. 
Let $\{\tau_j(\overline \omega)\}_{j\in \N}$ be the $I_\pm^\N$ valued process defined by (\ref{process}). Then, with the notation
\be
\mbox{\em Prob} ((\tau_n(\overline\omega), \dots, \tau_1(\overline\omega), \tau_0(\overline\omega))=(\tau_n, \dots, \tau_1, \tau_0))=T(\tau_n, \dots, \tau_1, \tau_0), \ \ \ \ n\in\N,
\ee
we have
\bea\label{ttau}
T(\tau_n, \dots, \tau_1, \tau_0)&=&|a_{\tau_0}|^2\sum_{\pi_1, \dots, \pi_n\in\Omega}p(\pi_1){\mathbb P}(\sigma_{r(\tau_{1})}(\pi_1),\pi_2)\cdots{\mathbb P}(\sigma_{r(\tau_{n-1})}(\pi_{n-1}),\pi_n)\times  \nonumber\\
&&\times \bra \tau_n | C(\pi_n)\tau_{n-1}\ket  
 \cdots  \bra \tau_1 | C(\pi_1)\tau_{0}\ket .
\eea
\end{lem}
\proof We start with $T(\tau_0)=|a_{\tau_0}|^2$, according to the initial condition, and 
\bea\nonumber
T(\tau_1, \tau_0)&=&|a_{\tau_0}|^2\mbox{Prob}(\omega_1 \ \mbox{s.t.} \ \sigma_{0}(\omega_1)(\tau_0)=\tau_1)\\
&=&|a_{\tau_0}|^2\sum_{\pi_1\in \Omega}\delta_{\tau_1, \sigma_{0}(\pi_1)(\tau_0)} p(\pi_1)=|a_{\tau_0}|^2\sum_{\pi_1\in \Omega}\bra\tau_1| C(\sigma_{0}(\pi_1))\ \tau_0\ket p(\pi_1).
\eea Note that since $\sigma_0$ is the identity, $T(\tau_1,\tau_0)=\E_p(\bra\tau_1| C(\omega)\ \tau_0\ket )|a_{\tau_0}|^2$.
Then
\bea
T(\tau_2, \tau_1, \tau_0)&=&|a_{\tau_0}|^2\mbox{Prob}((\omega_1, \omega_2) \ \mbox{s.t.}  \ \sigma_{0}(\omega_1)(\tau_0)=\tau_1 \, \mbox{and} \, \sigma_{r(\tau_1)}(\omega_2)(\tau_1)=\tau_2)\\
&=&|a_{\tau_0}|^2\sum_{\pi_1, \pi_2\in \Omega}\delta_{\tau_2, \sigma_{r(\tau_1)}(\pi_2)(\tau_1)} \delta_{\tau_1, \sigma_{0}(\pi_1)(\tau_0)} 
 p(\pi_1)\P(\pi_1, \pi_2)\nonumber \\  &=&|a_{\tau_0}|^2\sum_{\pi_1, \pi_2\in \Omega}\bra\tau_2| C(\sigma_{r(\tau_1)}(\pi_2))\ \tau_1\ket\bra\tau_1| C(\sigma_{0}(\pi_1))\ \tau_0\ket p(\pi_1)\P(\pi_1, \pi_2),\nonumber 
\eea
and, by induction
\bea\label{ttaup}
&&T(\tau_n, \dots, \tau_1, \tau_0)=|a_{\tau_0}|^2\sum_{\pi_1, \dots, \pi_n\in\Omega}p(\pi_1){\mathbb P}(\pi_1,\pi_2)\cdots{\mathbb P}(\pi_{n-1},\pi_n)\times \\ \nonumber
&&\hspace{.5cm}\times \bra \tau_n | C(\sigma_{\sum_{s=1}^{n-1}r(\tau_{s})}(\pi_n))\tau_{n-1}\ket  
 \cdots  \bra \tau_1 | C(\sigma_{0}(\pi_1)))\tau_{0}\ket .
\eea
Using the properties of $\sigma$, the measure invariant representation of $\Z^d$,  we get for any $j\geq 1$ with $\tilde \pi_j=\sigma_{\sum_{s=1}^{j-1}r(\tau_{s})}(\pi_j)$,
\be
\sum_{\pi_j\in\Omega}{\mathbb P}(\pi_{j-1},\pi_j)\bra \tau_j | C(\sigma_{\sum_{s=1}^{j-1}r(\tau_{s})}(\pi_j))\tau_{j-1}\ket  = \sum_{\tilde \pi_j\in \Omega}{\mathbb P}(\sigma_{r(\tau_{j-1})}(\tilde \pi_{j-1}),\tilde \pi_j)\bra \tau_j | C(\tilde\pi_j)\tau_{j-1}\ket,
\ee
which ends the proof.
\ep\\

The distribution of $\{\tau_j(\omega)\}_{j\in\N}$ is neither stationary, nor Markovian, in general. But we can express it in a more convenient way as follows.

\medskip

Consider the space $\C^{2d}\otimes\C^{|\Omega|}$ with orthonormal basis denoted by
$\{|\tau\otimes\pi\ket\}_{\tau\in I\pm, \pi\in \Omega}$. Let $ N\in M_{2d |\Omega|}(\R^+)$ be defined by its matrix elements 
\be\label{nnn}
\bra \tau'\otimes\pi'  | N \ \tau\otimes\pi \ket=\bra \tau' | C(\pi')\ \tau\ket {\mathbb P}(\sigma_{r(\tau)}(\pi) ,\pi')=\delta_{\tau', \pi'(\tau)}{\mathbb P}(\sigma_{r(\tau)}(\pi) ,\pi'),
\ee
and the vectors $\Psi_1=\sum_{\tau\in I_\pm \atop \pi\in \Omega}|\tau\otimes\pi \ket$ and $A(\tau_0)=\sum_{\pi,\tau}   |a_{\tau_0}|^2p(\pi)\bra\tau|C(\pi)\tau_0\ket  |\tau\otimes\pi \ket$.
Then, (\ref{ttau}) reads
\be
T(\tau_n, \dots, \tau_1, \tau_0)=\big\bra \Psi_1|(|\tau_{n}\ket\bra \tau_{n}|\otimes \I)N (|\tau_{n-1}\ket\bra \tau_{n-1}|\otimes \I)\dots N (|\tau_{1}\ket\bra \tau_{1}|\otimes \I)A(\tau_0)\big\ket.
\ee
Introducing also the matrices $D(y)$ and $N(y)$ on $\C^{2d}\otimes\C^{|\Omega|}$ with $y\in \T^d$ by
\be
D(y)=d(y)\otimes \I, \ \ \mbox{where} \ \ d(y)=\sum_{\tau\in I_\pm}e^{iyr(\tau)}|\tau\ket\bra \tau | \ \mbox{and} \ \ N(y)=D(y)N 
\ee
we can express the characteristic function $\Phi^T_n: \T^d\ra\C$ of the random variable $S_n(\overline \omega)=\sum_{j=1}^nr(\tau_j(\overline \omega))$ as
\bea
\Phi_n^T(y)&=&\sum_{\tau_n, \tau_{n-1}, \dots, \tau_0\in I_\pm}e^{iy\sum_{j=1}^nr(\tau_j)}T(\tau_n, \dots, \tau_1, \tau_0)\\ \nonumber
&=&\big\bra \Psi_1|(N(y))^{n-1}B(y) \big\ket,  \ \ \mbox{where} \ \ \ B(y)=D(y)\sum_{\tau_0\in I_\pm}A(\tau_0).
\eea

At this point, we can apply the same methods as above to describe the large $n$ behavior of $S_n(\overline\omega)$ by studying the asymptotic behavior of the suitably rescaled characteristic function $\Phi_n^T(y)$ under appropriate spectral assumptions on $N$. 

\medskip

Note that $N$ is a stochastic matrix and that $\P$ and $p$ are invariant under $\sigma_x$ so that 
we have
\be
N^T\Psi_1=\Psi_1 \ ,\ \ N\chi_1=\chi_1\ , \ \ \mbox{and }  \ \ \ \|N\|=\spr (N)=1
\ee 
with
\be \Psi_1=\sum_{\tau\in I_\pm \atop \pi\in \Omega}|\tau\otimes\pi\ket \ \ \mbox{and} \ \ \chi_1=\sum_{\tau\in I_\pm \atop \pi\in \Omega}p(\pi)|\tau\otimes\pi \ket.
\ee
Also, $D(y)$ being unitary for $y$ real, we have $\|N(y)\|\leq 1$ for all $y\in \T^d$.
\medskip \\
Assumption $\bf \tilde{\mbox{S}}$:
\be
\sigma(N)\cap D(0,1)= \{1\}, \ \ \mbox{and this eigenvalue is simple.} 
\ee 
\begin{rems} The corresponding spectral projector of $N$ reads $|\chi_1\ket\bra\Psi_1|/(2d).$
\\Again, it is enough to assume that $\bf \tilde{\mbox{S}}$ holds for the restriction of $N$ to the $N(y)^*$-cyclic subspace generated by $\Psi_1$.
\end{rems}

The perturbative arguments given in Section 4 leading to Corollary \ref{cf} by means of L\'evy Theorem apply here. For $y\in \C^d$ in a neighborhood of the origin, let $\lambda_1(y)$ be the simple analytic eigenvalue of $N(y)$ emanating from 1 at $y=0$. Let $\overline v \in \R^d$ and the non negative  matrix $\Sigma\in M_{d}(\R) $ defined by the expansion
\be
\lambda_1(y)=1+iy\overline v -\frac{1}{2}\bra y|\Sigma y\ket +O(\|y\|^3).
\ee
Explicit computations yield for any $y\in\C^d$
\bea\label{vbs}\nonumber
\overline v&=& \frac{1}{2d}\sum_{\tau\in I_\pm}r(\tau)\equiv \overline r\\
\bra y |\Sigma y\ket&=&-\frac1d \sum_{\tau\in I_\pm}\frac{(yr(\tau))^2}{2}
\\ \nonumber
& &-\frac{1}{d}\left(  \sum_{\tau, \tau'\in I_\pm} (yr(\tau))(yr(\tau'))\left\{
\bra \tau\otimes\eta_1| S(1) \tau'\otimes\eta_p\ket
-\frac{1}{2d}\right\} \right),
\eea
where $S(1)$ is the reduced resolvent of $N$ at $1$, $\eta_1=\sum_{\pi}|\pi\ket$ and $\eta_1=\sum_{\pi}p(\pi)|\pi\ket.$

\begin{prop}\label{prra} Let $\ffi_0=\sum_{\tau_0\in I_\pm}a_{\tau_0}|\tau_0\ket$ and let $S_n(\overline\omega )=\sum_{j=1}^nr(\tau_j(\overline\omega ))$, with $\tau_j(\overline\omega )$ defined by (\ref{process}).  Assume {\bf $\widetilde M$} and $\bf \tilde{\mbox{S}}$ and let $\Sigma$ be defined by (\ref{vbs}).
Then, 
if $\Sigma>0$ , we have as $n\ra\infty$
 \bea
\frac{{S_n}(\overline\omega )}{n}&{\cD \atop \longrightarrow }& \overline{r}\\
 \frac{S_n(\overline\omega)-n\overline v}{\sqrt n}&{\cD \atop \longrightarrow }&\ X^\omega\simeq \cN(0,\Sigma).
 \eea
\end{prop}

 As a consequence,  for any sample of random coin matrices, we obtain the following long time asymptotics of the quantum mechanical random probability distribution $W_\cdot^{\ffi_0}(n)$ of the variable $X_n^{\overline\omega}$, whose characteristic function is defined by (\ref{carran}).
 
\begin{thm}\label{randd} Under the assumptions of Proposition \ref{prra}, the following random variables converge in distribution as $n\ra\infty$:  \\
\bea
e^{-iy\overline{r}\sqrt n }\Phi_n^{\ffi_0}(y/\sqrt n)=\sum_{\tau_0\in I_\pm}|a_{\tau_0}|^2\left(e^{iy\frac{1}{\sqrt n} (S_n(\overline \omega) -n\overline{r} )}\right)\longrightarrow e^{iyX^\omega} , 
\eea
where $X^\omega\simeq \cN(0,\Sigma)$. Moreover, for any $(i,j)\in \{1, 2, \dots, d\}^2$, as $n\ra\infty$, we have in distribution,
\bea
&&\frac{\bra X_i\ket^{\overline\omega}_{\psi_0}(n)}{n}\overline{r}_i \\
&&\frac{\bra (X-n\overline{r})_i (X-n\overline{r})_j \ket^{\overline\omega}_{\psi_0}(n)}{n}
\longrightarrow \D_{jk}^{\omega}
\eea
where $\D_{jk}^\omega$ is distributed according to the law of $X_j^\omega X_k^\omega$, where 
$X^\omega\simeq \cN(0,\Sigma)$.

 \end{thm}
\proof Identical to that of Corollary 6.8 in \cite{J1}.

\bigskip

\subsection{Specific Case}

Let us close this section by providing an example that satisfies the assumption $\bf \tilde{\mbox{S}}$. It is the case where the kernel $\P$ depends on the second index only, i.e., when the permutations $\{\omega(j)\}_{j\in\N}$ are i.i.d. and distributed according to $p$.

\begin{prop}
Assume {$\widetilde{\bf  M}$} with a kernel $\P$ satisfying
$
\P(\pi',\pi)=p(\pi).
$  
Let $P$ be the bi-stochastic matrix acting on $\C^{2d}$ defined by 
\be\label{bistoc}
P=\sum_{\pi\in\Omega}p(\pi)C^T(\pi)\equiv\E_p( C^T(\omega))
\ee
and assume it irreducible and aperiodic. Then $\bf \tilde{\mbox{S}}$ holds and Theorem \ref{randd} applies with $\Sigma$ given by
\bea\label{covma}
\Sigma_{ij}&=&-\frac{1}{2d}\bra r_i |r_j\ket+\overline r_i \overline r_j-\frac{1}{2d}\left(
 \bra r_i | S(1)r_j \ket+  \bra r_j | S(1)r_i \ket\right ),
\eea
with $S(1)$ the reduced resolvent of $P$ at $1$ and, for $j\in\{1,\dots,d\}$,  $r_j=\sum_{\tau\in I_\pm}r_j(\tau)|\tau\ket\in \C^{2d}$.
\end{prop}
\proof
 In this case, (\ref{nnn}) reduces to
 \be
 \bra \tau'\otimes\pi'  | N \ \tau\otimes\pi \ket=\bra \tau' | C(\pi')\ \tau\ket p(\pi'),
\ee
so that we can write with $\eta_1=\sum_{\pi}|\pi\ket$,
\be
N^T=\sum_\pi p(\pi)C^T(\pi)\otimes |\eta_1\ket\bra\pi |.
\ee
Accordingly, for any $\xi\in \C^{2d}\otimes \C^{|\Omega|}$, we have
\be
N^T\xi=\zeta(\xi)\otimes|\eta_1\ket, \ \ \mbox{with } \ \ \zeta(\xi)=\sum_{\pi\in\Omega}p(\pi)C^T(\pi)\bra\pi |\xi\ket_{\C^{|\Omega|}}, \ \  \bra\pi|\xi\ket_{\C^{|\Omega|}}\in \C^{2d}.
\ee
Hence, any eigenvector $\tilde \Psi$ with eigenvalue $e^{i\theta}$, $\theta\in\R$, needs to be of the form $\tilde \Psi=\psi\otimes \eta_1$ with
\be\label{evuc}
e^{i\theta}\psi=\sum_{\pi\in\Omega}p(\pi)C^T(\pi)\psi=P\psi.
\ee
The matrix $P$ being bi-stochastic, irreducible and aperiodic, there exists only one solution to (\ref{evuc}), given by $\psi=\sum_{\tau\in I_\pm}|\tau\ket$ and $e^{i\theta}=1$, which shows that $\bf \tilde{\mbox{S}}$ holds.\\

The expectation $\overline v$ and correlation matrix $\Sigma$ can be obtained from Theorem 6.6 in \cite{J1}. Indeed, under our assumptions,  Lemma \ref{nonstat} shows that  the process $\{\tau_j(\overline \omega)\}_{j=1,\dots, n}$ is a Markov chain on $I_\pm$, with kernel $P=\E_p( C^T(\omega)) $ and initial distribution $p_0(\tau_0)=|a_{\tau_0}|^2$:
\bea
T(\tau_n, \dots, \tau_1, \tau_0)&=&|a_{\tau_0}|^2\sum_{\pi_1, \dots, \pi_n\in\Omega}p(\pi_1)p(\pi_2)\cdots, p(\pi_n) \bra \tau_n | C(\pi_n)\tau_{n-1}\ket \cdots  \bra \tau_1 | C(\pi_1)\tau_{0}\ket  \nonumber\\
&=&P(\tau_n,\tau_{n-1})\cdots P(\tau_1,\tau_0)p_0(\tau_0).
\eea
 The aforementioned result provides the characteristics $\overline v$ and $\Sigma$ (\ref{covma}) of the functional central limit theorem for the Markov chain $\{\tau_j(\overline \omega)\}_{j=1,\dots, n}$ corresponding to the random variable $S_n(\overline\omega)=\sum_{j=1}^nr(\tau_j(\overline\omega))$. 
\ep\\
\proof
With (\ref{simple}), (\ref{ttau}) reads
\bea\label{ttausimple}
T(\tau_n, \dots, \tau_1, \tau_0)&=&|a_{\tau_0}|^2\sum_{\pi_1, \dots, \pi_n\in\Omega}p(\pi_1)p(\pi_2)\cdots, p(\pi_n)\times \\ \nonumber
&&\times \bra \tau_n | C(\sigma_{\sum_{s=1}^{n-1}r(\tau_{s})}(\pi_n))\tau_{n-1}\ket  
 \cdots  \bra \tau_1 | C(\sigma_{0}(\pi_1)))\tau_{0}\ket,
\eea
where for all $j\geq 1$, thanks to the fact that $\sigma_x$ is measure preserving, 
\be
\sum_{\pi_j}p(\pi_j)\bra \tau_j | C(\sigma_{\sum_{s=1}^{j-1}r(\tau_{s})}(\pi_j))\tau_{j-1}\ket =\sum_{\pi_j}p(\pi_j)\bra \tau_j | C(\pi_j))\tau_{j-1}\ket.
\ee
Setting $P(\tau',\tau)=\E_p(\bra \tau'|C^T(\omega)\tau\ket)$ and $p_0(\tau)=|a_{\tau}|^2$, we can write
\be
T(\tau_n, \dots, \tau_1, \tau_0)=P(\tau_n,\tau_{n-1})\cdots P(\tau_1,\tau_0)p_0(\tau_0),
\ee
which proves the claim.
\ep
\begin{rem}
Actually, a strong law of large numbers holds in this case, i.e. $\lim_{n\ra\infty}S_n(\overline\omega)/n\ra \overline r$ almost surely.
\end{rem}

\section{Uncorrelated example}\label{uncorr}

In this last section, we briefly present two cases where the random coin matrices are chosen in an uncorrelated way, in order to complete the picture. In a sense, it can be viewed as the limiting case where the representation $\sigma$ of $\Z^d$ is such that the periodicity lattice $\Gamma$ is infinite. This is the complete opposite situation to the one considered in  \cite{J1}, where all coin matrices where identical in space, at all time step. Nevertheless, the methods developed in that paper apply here too.

\bigskip

We recall some notations used in Section 2.1 in \cite{J1}:
let $x_s=\sum_{j=1}^{s-1}r(\tau_j)$, $x'_j=\sum_{j=1}^{s-1}r(\tau'_j)$, then the generic term in Lemma \ref{wkn} reads 
\bea
\bra \tau_{s-1}'|C^*_s(x'_s)\ \tau_s'\ket\bra \tau_s|C_s(x_s) \ \tau_{s-1}\ket = \overline{\bra \tau_{s}'|C_s(x'_s)\ \tau_{s-1}'\ket}\bra \tau_s|C_s(x_s) \ \tau_{s-1}\ket
\\\equiv\bra \tau_s\otimes\tau'_{s} | (C_s(x_s)\otimes \overline C_s(x'_s)) \ \tau_{s-1}\otimes\tau'_{s-1} \ket.
\eea
Let us introduce the unitary tensor product   
\be\label{tpv}
V_s(x,y)\equiv C_s(x)\otimes \overline C_s(y) \ \ \mbox{in}\ \C^{2d}\otimes \C^{2d}.
\ee 
Now consider the set of paths $G_n(K)$ in $\Z^{2d}$ from the origin to $K=\begin{pmatrix}k \\ k' \end{pmatrix}\in \Z^{2d}$ via the (extended) jump function defined by  
\be R: I_\pm^2\ra \Z^{2d}, \ \ \ R\begin{pmatrix}\tau_s \\ \tau_s' \end{pmatrix}=\begin{pmatrix}r(\tau_s) \\ r(\tau_s') \end{pmatrix},
\ee that is paths of the form $(T_1, \cdots,  T_{n-1}, T_n)$, where $T_s=\begin{pmatrix}\tau_s \\ \tau_s' \end{pmatrix}\in I_\pm^2$, $s=1,2, \dots,n$, and $ \sum_{s=1}^n R(T_s)=K$.
For $s\geq 2$ let $X_s=\sum_{j=1}^{s-1} R(T_j)$, while $X_1=0$. This last condition states that we start the walk at the origin.

With these notations, we consider the {\it complex weight} of $n$-step paths in $\Z^{2d}$ from the origin to $K$, with last step $T$, defined by
\be
W_K^T(n)=\sum_{(T_1, \cdots,  T_{n-1})\in {I^2_\pm}^{n-1} \ \mbox{\tiny s.t.} \atop {(T_1, \cdots,  T_{n-1}, T)\in G_n(K)}}\bra T|V_n(X_n)T_{n-1}\ket\cdots\bra T_2| V_2(X_2) T_1\ket \bra T_1 V_1(0)\chi_0\ket,
\ee
with $\chi_0$ defined via the decomposition
\be
\ffi_0=\sum_{\tau\in I_\pm}a_\tau |\tau\ket \ \Rightarrow \ \chi_0=\ffi_0\otimes\overline{\ffi_0}=\sum_{(\tau, \tau')\in I^2_\pm}a_\tau \overline a_{\tau'}|\tau\otimes \tau'\ket.
\ee
The expectation of this complex weight is the key quantity to analyze the averaged characteristic function (\ref{averex}), see \cite{J1}.
Under certain  assumptions on the distributions of the matrices $C_j(x)\in \Omega$, $\Omega$ finite, for simplicity, some cases can be readily studied using this method. 

\medskip 

\noindent{\bf Assumption A:} 

\noindent a) The matrices $V^\omega_j(X)$ are distributed so that 
\bea
\P(V^\omega_n(X_n)=Z_n, V^\omega_{n-1}(X_{n-1})=Z_{n-1},\cdots, V^\omega_1(X_0)=Z_0)=\prod_{j=1}^n \P(V^\omega_j(X_j)=Z_j).
\eea 
b) The expectation $\E(V^\omega_k(X))$ is independent of the position $X$:
$$Q_k= \sum_{Z\in\Omega\otimes\Omega}Z\P(V^\omega_k(X)=Z) =\E(V^\omega_k(X)).$$

\medskip
Assumption A is clearly satisfied in the following cases:\\

{\bf Case 1:}
Assuming that the distributions of the matrices $C_s(x)$ are i.i.d in time and position, requirement a) is satisfied with $\P(V^\omega_j(X)=Z)$ independent of $j$. Moreover, $\P(V^\omega(x,y)=Z)=\P_{O}(Z)$, for all  $x\neq y$,  and $\P(V^\omega(x,x)=Z)=\P_{D}(Z)$, for all $x$. Further assuming 
\be
\sum_{Z\in\Omega\otimes\Omega}Z\P_{O}(Z)=\sum_{Z\in\Omega\otimes\Omega}Z\P_{D}(Z)\equiv Q,
\ee
we meet requirement b) as well.

\medskip

 {\bf Case 2:} The following holds:\\
i) For $X\in\Z^{2d}$,  $V^\omega_j(X)$ is a Markov chain in time on $\Omega\otimes \Omega$ with initial distribution $p_X$ and transition matrix $\P_X$. While  for $X\neq Y$, the random variables $V^\omega(X), V^\omega(Y)$ are independent.   
\\ ii)  The jump function $R: I_\pm^2\ra \Z^{2d}$ is one to one and any $X\in\Z^{2d}$  can only be reached at most once on $\{\sum_{T\in I^2_\pm} \alpha_T R(T), \alpha_T \in\N\}\subset \Z^{2d}$ along any path $X_s=\sum_{j=1}^{s-1} R(T_j)$, $s\in\N$.
\\ iii) $\E(V_j^{\omega}(X))=\sum_{Z\in\Omega\otimes\Omega}Z \bra p_X| \P_X^{j-1}Z\ket\equiv Q_j$ is independent of $X$, for any $j\in\N$.

\bigskip
Under assumption A, we get the following expression for the expectation of $W_K^T(n)$
\bea
\E(W_K^T(n))= \sum_{(T_1, \cdots,  T_{n-1})\in {I^2_\pm}^{n-1} \ \mbox{\tiny s.t.} \atop {(T_1, \cdots,  T_{n-1}, T)\in G_n(K)}}\bra T|Q_n T_{n-1}\ket\prod_{j=2}^{n-1}\bra T_j|Q_j T_{j-1}\ket \bra T_1|Q_1\chi_0\ket.
\eea
\medskip
Now we proceed as in \cite{J1}.
Introduce the vectors in $\C^{4d^2}\simeq \C^{2d}\otimes \C^{2d}$ with $Y\in \T^{2d} $ and $n\geq 0$
\be
 {\bf \Phi}_n(Y) = \sum_{T\in I_\pm^2} \sum_{K\in\Z^{2d}}e^{iYK}W_K^T(n)\, |T\ket \ \ \mbox{and }\ {\bf \Phi}_0 =  \sum_{T\in I_\pm^2} A_T\, |T\ket.
\ee

 Using the notation 
\be D(Y)=\sum_{T\in I^2_\pm} e^{iYR(T)} \, |T \ket \bra T |,  \ \ \mbox{with }\ Y\in \T^{2d} \  \mbox{and }\  M_k(Y)=D(Y)Q_k,
\ee
we obtain the following expression for the expectation
\bea
\E({\bf \Phi}_n(Y))&=&M_n(Y)M_{n-1}(Y)\cdots M_1(Y){\bf \Phi}_0.
\eea
 We get the following expression for the expectation of characteristic function (Proposition 2.9 in \cite{J1})
\be
\E(\Phi^{\ffi_0}_n(y))
=\int_{\T^d}\bra{\bf \Psi}_1 |M_n(Y_v)M_{n-1}(Y_v)\cdots M_1(Y_v){\bf \Phi}_0\ket \, d{\tilde v}, 
\ee  where
\bea
{\bf \Psi}_1=\sum_{T\in H_\pm} |T\ket=\sum_{\tau\in I_\pm}|\tau\otimes\tau\ket \ \ \mbox{and }\ Y_v=\begin{pmatrix}y-v\\ v \end{pmatrix}\in \R^{2d}
\eea

At this stage the exact dependence of the matrix $M_j$ on time $j$ becomes crucial.  In Case~1, $M_j=M$ for all $j$, so that we are directly lead to the asymptotic study of 
\be
\int_{\T^d}\bra{\bf \Psi}_1 |M^n(Y_v){\bf \Phi}_0\ket \, d{\tilde v},
\ee
as in \cite{J1}, which allows us to get diffusion properties and deviation estimates as in sections \ref{diff}, \ref{moder}, \ref{LargeDev}, provided $M(Y_v)$ satisfies the required spectral properties.

In order to deal with Case 2 for non stationary initial distribution $p_X$, an analysis of the large $j$ behavior of $Q_j$ based on the spectral properties of the transition matrix $\P_X$ is in order. This should provide the necessary information to reach similar conclusions as in Case 1.

\end{document}